\documentclass[12pt,draftclsnofoot,onecolumn]{IEEEtran}
\usepackage{graphicx}
\usepackage{bbm}
\usepackage{booktabs}
\usepackage{multirow}
\usepackage{amsthm} 
\usepackage{lipsum}
\usepackage{mathtools}
\usepackage{cuted}
\usepackage{amsmath}
\usepackage[cmintegrals]{newtxmath}
\usepackage{bm}
\usepackage{algorithmic}
\usepackage{epstopdf}
\usepackage{array}
\usepackage{algorithm}
\usepackage{color}
\usepackage{makecell}
\usepackage{tabularx} 
\usepackage{cite}
\usepackage{multirow}
\usepackage{multicol}
\usepackage{enumerate}
\usepackage{epsfig}
\usepackage[tight]{subfigure}

\begin{document}
    \title{Multi-Modal Variable-Rate CSI Reconstruction for FDD Massive MIMO Systems}

\author{Yunseo~Nam, Jiwook~Choi, and Saewoong~Bahk, \IEEEmembership{Senior Member, IEEE}
\thanks{Yunseo Nam and Saewoong Bahk are with the Department of Electrical and Computer Engineering, INMC, Seoul National University, Seoul 08826, South Korea  (e-mail: ysnam@netlab.snu.ac.kr; sbahk@snu.ac.kr).}
\thanks{Jiwook Choi is with the Korea Atomic Energy Research Institute (KAERI), Daejeon 34057, Republic of Korea (e-mail: jiwook@kaeri.re.kr).}
}

\maketitle

\setlength\arraycolsep{2pt}
\newcommand{\argmax}{\operatornamewithlimits{argmax}}
\newcommand{\argmin}{\operatornamewithlimits{argmin}}
\makeatletter
\newcommand{\vast}{\bBigg@{3.5}}
\newcommand{\Vast}{\bBigg@{4.5}}
\makeatother

\begin{abstract} 

In frequency division duplex (FDD) systems, acquiring channel state information (CSI) at the base station (BS) traditionally relies on limited feedback from mobile terminals (MTs). However, the accuracy of channel reconstruction from feedback CSI is inherently constrained by the rate-distortion trade-off. To overcome this limitation, we propose a multi-modal channel reconstruction framework that leverages auxiliary data, such as RGB images or uplink CSI, collected at the BS. By integrating contextual information from these modalities, we mitigate CSI distortions caused by noise, compression, and quantization. At its core, we utilize an autoencoder network capable of generating variable-length CSI, tailored for rate-adaptive multi-modal channel reconstruction. By augmenting the foundational autoencoder network using a transfer learning-based multi-modal fusion strategy, we enable accurate channel reconstruction in both single-modal and multi-modal scenarios. To train and evaluate the network under diverse and realistic wireless conditions, we construct a synthetic dataset that pairs wireless channel data with sensor data through 3D modeling and ray tracing. Simulation results demonstrate that the proposed framework achieves near-optimal beamforming gains in 5G New Radio (5G NR)-compliant scenarios, highlighting the potential of sensor data integration to improve CSI reconstruction accuracy.

\end{abstract}
\begin{IEEEkeywords}
Massive multiple-input multiple-output, channel state information feedback, deep neural network, multi-modal learning.
\end{IEEEkeywords}

\section{Introduction}
To meet the growing demand for high data rates, modern wireless systems utilize the abundant frequency spectrum available in the millimeter-wave bands (24GHz$\sim$71GHz) \cite{Swindlehurst, Boccardi}. At these frequencies, beamforming enabled by massive multiple-input multiple-output (MIMO) antenna arrays is essential to counteract severe signal attenuation \cite{Larsson, Ayach, Han}. To fully harness the benefits of beamforming, base stations (BSs) require accurate and instantaneous channel state information (CSI). However, acquiring precise CSI is challenging due to the large number of antennas and subcarriers, which greatly increase the size of the channel matrix. This problem is particularly pronounced in frequency division duplex (FDD) systems, where the separation between uplink and downlink frequencies prevents direct downlink channel estimation from uplink reference signals. In FDD systems, BSs generally depend on feedback from mobile terminals (MTs) to obtain CSI \cite{Love, Guo2}. Unfortunately, the channel estimated at the MT is often compromised by noise and the limited number of downlink reference signals. Moreover, the application of aggressive channel compression and quantization to reduce feedback overhead further deteriorates the quality of the CSI. Consequently, the imperfect CSI received at the BS significantly impairs the ability to achieve high beamforming gains.

To alleviate beamforming gain degradation due to limited CSI feedback, various methods have been developed to derive compact, discrete representations of the channel. Traditional approaches represent the channel using a few physical ray parameters, such as amplitude, angle, phase, and delay \cite{Shen, Ju1, 3GPP1, Qin}. For instance, the Type II precoding matrix indicator (PMI) codebook in the 5th generation new radio (5G NR) standard captures ray directions shared across the entire frequency band, along with the amplitudes and phases for each frequency subband \cite{3GPP1, Qin}. However, as the number of subcarriers increases, feedback overhead scales linearly, leading to redundancy due to the strong correlation among amplitude-phase pairs across subbands. To address this inefficiency, compressed sensing (CS) techniques have been widely adopted \cite{Eltayeb, Huang, Gao, Kulsoom}. By exploiting the sparsity of high frequency channels in the angular-delay domain, CS-based CSI feedback can maintain high reconstruction accuracy even under significant compression. However, CS-based methods typically rely on computationally intensive iterative algorithms, which pose challenges for real-world deployment.

\begin{figure}[t]
	\centering
  \includegraphics[width=0.6\textwidth]{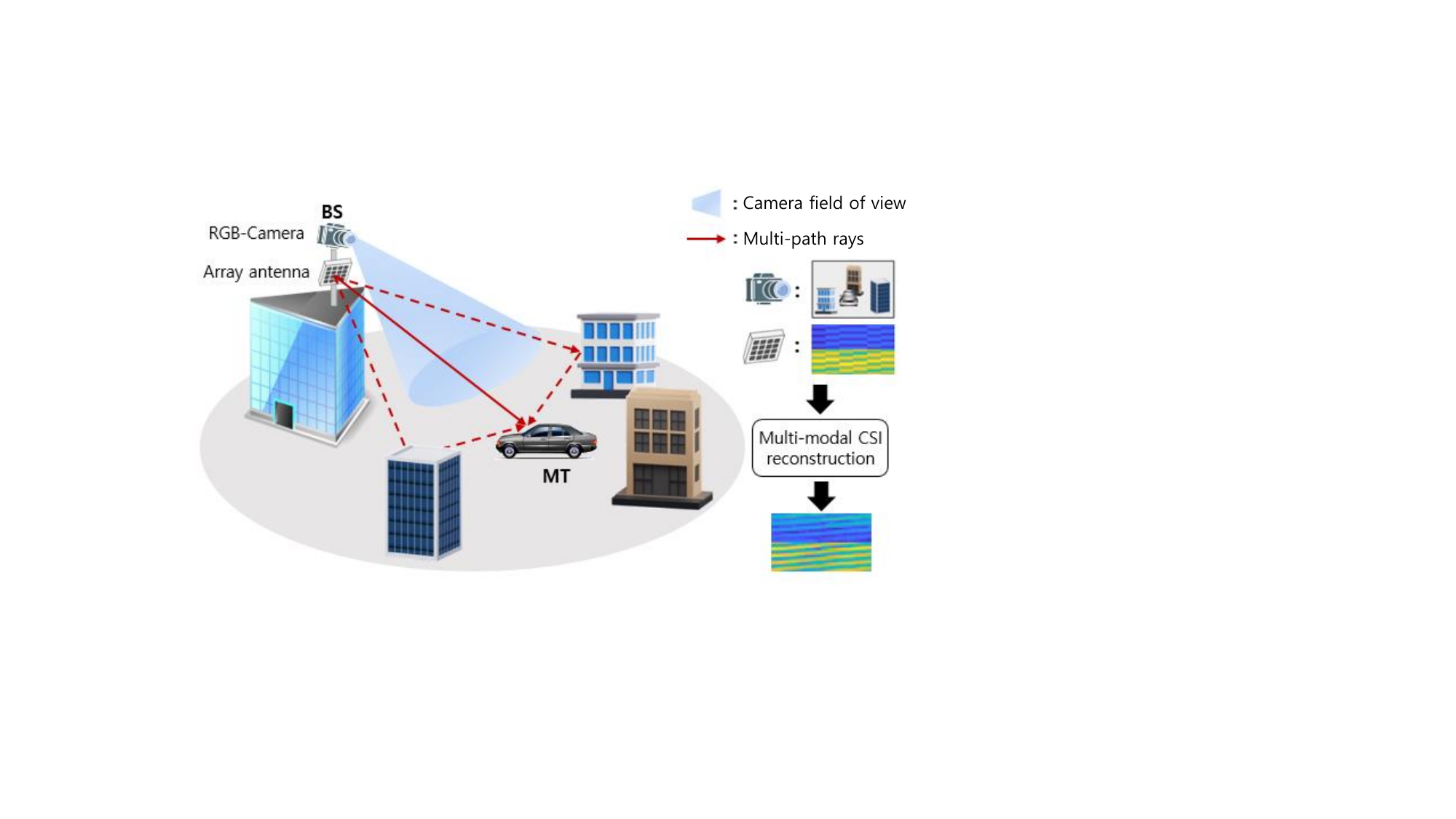}
  \caption{Multi-modal channel reconstruction using image data and wireless data.} \label{Fig6}
\end{figure}

Deep learning (DL)-based CSI feedback mechanisms \cite{3GPPAI, Liang, Wen, Mashhadi, Lu, Guo, Liang2, Jeon1, Liu, Zeng} have emerged as a promising solution for fast and accurate CSI compression and reconstruction in 5G-Advanced (3rd Generation Partnership Project (3GPP) Rel. 18 \cite{3GPPAI}). Despite their advantages, the static architecture of neural networks limits their flexibility for real-time adaptation. For practical deployment, DL-based feedback techniques must support variable-length CSI bit streams to accommodate diverse reporting configurations \cite{Guo, Liang2, Jeon1}. Some architectures address this challenge by employing network layers that progressively downsample the CSI \cite{Guo}. However, this approach increases the network size in proportion to the number of supported rates, leading to scalability issues. To enable variable-rate feedback without adding network complexity, one approach directly discards the least significant bits of CSI \cite{Liang2}. But, discarding CSI bits without modifying the quantization rules can impair the network's ability to learn efficient variable-rate channel representations. A more sophisticated technique involves nested vector quantization (VQ), which selects low rate child codebooks from a high rate parent codebook \cite{Jeon1}. While effective, this approach requires the BSs and MTs to share all possible nesting configurations. Additionally, VQ-based methods demand extensive training parameters and computational resources, which can be burdensome for MTs with limited memory and power capabilities. These limitations highlight the need for DL-based feedback mechanisms that can achieve variable-rate feedback with minimal computational overhead and enhanced adaptability.

Achieving high channel reconstruction accuracy with minimal feedback overhead has been a central goal of CSI feedback research \cite{Shen, Ju1, 3GPP1, 3GPPAI, Qin, Eltayeb, Huang, Gao, Kulsoom, Liang, Wen, Mashhadi, Lu, Guo, Liang2, Jeon1, Liu, Zeng}. However, the reconstruction accuracy achievable with limited CSI feedback is fundamentally constrained by the rate-distortion trade-off \cite{Inf}. A promising strategy to address this limitation involves leveraging sensor data from diverse modalities \cite{FLiu, Zheng}. Unlike downlink CSI, sensor data, such as RGB images or uplink CSI offer high resolution and are less affected by noise, making them valuable for enhancing channel reconstruction. Among these modalities, deploying camera sensors at the BS is particularly advantageous due to the dominance of the line-of-sight (LOS) path at high carrier frequencies \cite{Zheng, Kim, Ahn, Ahn2}. Modern cameras provide hundreds of millions of pixels at relatively low deployment costs, enabling neural networks to extract environment-aware context for tasks such as MT positioning \cite{Kim}, handover \cite{Ahn}, and beamforming \cite{Ahn2}. Despite these advantages, camera sensors have inherent limitations. They are ineffective in non-line-of-sight (NLOS) scenarios and are sensitive to adverse weather conditions. Furthermore, visual data alone cannot fully characterize the channel, as it lacks critical parameters such as MT array orientation and Doppler shift. As such, relying exclusively on sensor data during channel reconstruction introduces inherent blind spots.

To leverage the advantages of sensor data while maintaining the reliability of conventional CSI feedback methods, we propose a hybrid approach that incorporates sensor data to enhance the channel reconstruction process. The core idea is to extract supplementary information from sensor data originating in the same physical environment as the downlink CSI \cite{Shimomura}. In this approach, neural networks learn positional information (e.g., the locations of MTs and dominant channel clusters) from sensor data, while structural information about the channel matrix (e.g., frequency and spatial selectivity) is derived from wireless data. By effectively fusing information from these heterogeneous modalities, the network enhances channel reconstruction accuracy. A major challenge in this process is the lack of large paired datasets containing sensor and wireless channel data \cite{ViWi}. Acquiring real-world channel measurements is not only time-consuming but also requires expensive hardware, such as radio frequency equipment and software-defined radios. Moreover, synchronizing channel data with sensor data, which are captured using entirely different hardware systems, adds further complexity. To address this, we generate a synthetic dataset using 3D rendering tools and ray tracing simulators. This scalable and noise-free approach facilitates the development of a robust hybrid CSI reconstruction framework. The main contributions of this work are as follows:
\begin{itemize}
\item We propose a super-resolution channel reconstruction technique that jointly leverages sensor data (e.g., RGB image and uplink CSI) collected at the BS and CSI fed back from the MT. At its core, we develop an autoencoder in which the encoder, quantizer, and decoder collaboratively learn variable-rate binary channel representations. Our design allows the generation of CSI with arbitrary lengths using only a few hundred parameters and minimal computational overhead. A key feature of the quantization network is the generation disjoint feature vector spaces for different rates. This enables the multi-modal fusion network to autonomously and adaptively assess the significance of sensor data during variable-rate channel reconstruction.

\item We synthesize a wireless channel-sensor data paired dataset to train the hybrid channel reconstruction framework. We emulate real-world communication scenarios by modeling high quality 3D objects, performing ray-tracing simulations, and generating clustered delay line MIMO channels compliant with the 5G NR standard. Using these datasets, the neural networks are trained based on a transfer learning strategy. By building the multi-modal channel reconstruction mechanism upon a foundational variable-rate autoencoder, the framework allows the BS to operate in either a wireless-only mode or a sensor data-assisted mode.

\item We evaluate the multi-modal, variable-rate channel reconstruction network under 5G NR compliant simulation settings \cite{3GPP1,3GPP2,3GPP3}. We demonstrate that the variable-rate autoencoder achieves much higher beamforming gains compared to the Type II PMI codebook. We also demonstrate that the multi-modal fusion improves the beamforming gains by a significant margin.
\end{itemize}

 The rest of this paper is organized as follows. In Section II, we describe the massive MIMO system model and briefly explain the DL-assisted CSI feedback process. In Section III, we present the network architecture for multi-modal, variable-rate CSI reconstruction. In Section IV, we discuss the dataset generation method and the training method for our network. Simulation results are provided in V. Then, we conclude the paper in Section VI.

\textit{Notations: }Throughout this paper, vectors and matrices are denoted by bold lowercase and uppercase letters, respectively. The operators $\|\cdot\|_2$ and $\|\cdot\|_{\rm F}$ denote the Euclidean norm and the Frobenius norm, respectively. The transpose and the conjugate transpose of matrices are denoted by $(\cdot)^{\top}$ and $(\cdot)^{\rm H}$, respectively. For a random variable, we use $\mathbb{E}[\cdot]$ to denote the expected value. The real part of a complex number is denoted by ${\rm Re}\{\cdot\}$. For a real number, $\lfloor{\cdot} \rfloor$ denotes the floor function. We use ${\sf sg}(\cdot)$ to denote the stop-gradient operator that ignores gradient computation during backward computation.

\section{System Model}
In this section, we describe the massive MIMO-orthogonal frequency division multiplexing (OFDM) downlink system model and the CSI feedback mechanism using DL.

\subsection{Massive MIMO-OFDM Downlink}
We consider a single-cell massive MIMO system where a BS equipped with $N_{\rm t}\in\mathbb{N}$ transmit antennas serves a MT equipped with a single receive antenna. The system adopts $N_{\rm s}=12N_\text{RB}$ subcarriers, where $N_\text{RB}\in\mathbb{N}$ deontes the number of downlink resource blocks (RBs). Let the spatial-frequency domain channel matrix be ${\bf H}=[{\bf h}_1, {\bf h}_2,\ldots,{\bf h}_{N_{\rm s}}]\in{\mathbb C}^{N_{\rm t}\times N_{\rm s}}$, where ${\bf h}_n\in{\mathbb C}^{N_{\rm t}}$ is the channel vector of the $n$th subcarrier. The downlink transmit grid is denoted as ${\bf X}=[{\bf x}_1, {\bf x}_2,\ldots,{\bf x}_{N_{\rm s}}]\in{\mathbb C}^{N_{\rm t}\times N_{\rm s}}$. Then, the received signal ${y}_n\in\mathbb{C}$ at the $n$th subcarrier is given by
\begin{align}
    y_n = {\bf h}_n^{\rm H}{\bf x}_n+v_n,
\end{align}
where $v_n\in {\mathbb C}$ represents the additive noise. At high carrier frequencies, beamforming is essential to compensate for the high signal attenuation. Specifically, the $n$th subcarrier symbol $s_n\in\mathbb{C}$ is mapped to ${\bf x}_n$ as ${\bf x}_n={\bf p}_n s_n$, where ${\bf p}_n\in{\mathbb C}^{N_{\rm t}}$ is the frequency-dependent beamforming vector with unit norm ($\|{\bf p}_n\|_2=1$). With beamforming, the input-output relation between $s_n$ and $y_n$ becomes
\begin{align}
     y_n = {\bf h}_n^{\rm H}{\bf p}_n s_n+v_n.
\end{align}
Consequently, the downlink channel throughput $R$ is expressed as
\begin{align}
    R = \sum_{n=1}^{N_{\rm s}} \log_2\left(1+\|{\bf h}_n^{\rm H}{\bf p}_n\|_2^2\frac{\mathbb{E}[\|s_n\|_2^2]}{\mathbb{E}[\|v_n\|_2^2]}\right). \label{throughput}
\end{align}
To maximize $R$, each beamforming vector should be chosen to maximize the precoded channel gain $\|{\bf h}_n^{\rm H}{\bf p}_n\|_2$.

\subsection{DL-Assisted CSI Feedback Process}
In FDD MIMO systems, the quality of the CSI fed back to the BS directly influences downlink throughput. To obtain this CSI, the MT first estimates $\tilde{\bf H}\in{\mathbb C}^{N_{\rm t}\times N_{\rm s}}$ by receiving a dedicated reference grid from the BS. Once $\tilde{\bf H}$ is acquired, the MT performs channel compression and quantization to reduce feedback overhead:
\begin{align}
{\bf s}=f_{\text E}(\tilde{\bf H},B)\in\{0,1\}^B,
\end{align}
where $B\in\mathbb{N}$ is the target length of the CSI bit stream. This bit stream is then fed back to the BS via the uplink control channel. Using ${\bf s}$, the BS reconstructs the channel and generates the beamforming matrix:
\begin{align}
    \hat{\bf H}=f_{\text D}({\bf s})\in\mathbb{C}^{N_{\rm t}\times N_{\rm s}}, \nonumber\\
    {\bf P} = f_\text{P}(\hat{\bf H})\in\mathbb{C}^{ N_{\rm t} \times N_{\rm s}}.
\end{align}
Here, the functions $f_{\text{E}}(\cdot,B)$ and $f_{\text{D}}(\cdot)$ are implemented using neural networks and are trained to minimize the loss function
\begin{align}
L({\bf H}, \hat{\bf H}) =
    \bigg\|\frac{{\bf H}}{\|{\bf H}\|_{\rm F}}-\frac{\hat{\bf H}}{\|\hat{\bf H}\|_{\rm F}} \bigg\|^2_{\rm F}. \label{lossfunction}
\end{align}
Minimizing $L({\bf H}, \hat{\bf H})$ is equivalent to maximizing the averaged cosine similarity between the columns of ${\bf H}=[{\bf h}_1, {\bf h}_2,\ldots,{\bf h}_{N_{\rm s}}]$ and $\hat{\bf H}=[\hat{\bf h}_{1}, \hat{\bf h}_{2},\ldots,\hat{\bf h}_{N_{\rm s}}]$, i.e.,
\begin{align}
&\arg_{f_{\text E},f_{\text D}}\min L({\bf H}, \hat{\bf H}) \nonumber\\
& =\arg_{f_{\text E},f_{\text D}}\max \frac{1}{N_{\rm s}}\sum_{n=1}^{N_{\rm s}} \frac{{\rm Re}\left\{\big({\bf h}_n\big)^{\rm H} \big(\hat{\bf h}_{n}\big)\right\}}{\big\|{\bf h}_n\big\|_2  \big\| \hat{\bf h}_{n}\big\|_2}.
\end{align}
By normalizing the magnitudes of both ${\bf H}$ and $\hat{\bf H}$, the network focuses on maximizing normalized beamforming gains. Finally, the beamforming matrix is obtained by normalizing each reconstructed channel vector:
\begin{align}
{\bf P}=f_{\text P}(\hat{\bf H})=\Big[\frac{\hat{\bf h}_1}{\|\hat{\bf h}_1\|_2}, \frac{\hat{\bf h}_2}{\|\hat{\bf h}_2\|_2}, \ldots, \frac{\hat{\bf h}_{N_{\rm s}}}{\|\hat{\bf h}_{N_{\rm s}}\|_2}\Big].
\end{align}
Thus, the CSI bit stream ${\bf s}$ is mapped one-to-one onto the beamforming matrix ${\bf P}$, omitting the channel magnitude. This design parallels the role of the PMI in the 5G New Radio (NR) standard \cite{3GPP1}. Meanwhile, additional channel magnitude metric (e.g., the channel quality indicator) can be fed back separately to support modulation/coding scheme selection.

\section{Network Architecture}
In this section, we propose a DL-assisted multi-modal, variable-rate channel reconstruction framework. We illustrate the overall network flow in Fig. 2.

\begin{figure*}[t]
	\centering
  \includegraphics[width=0.9\textwidth]{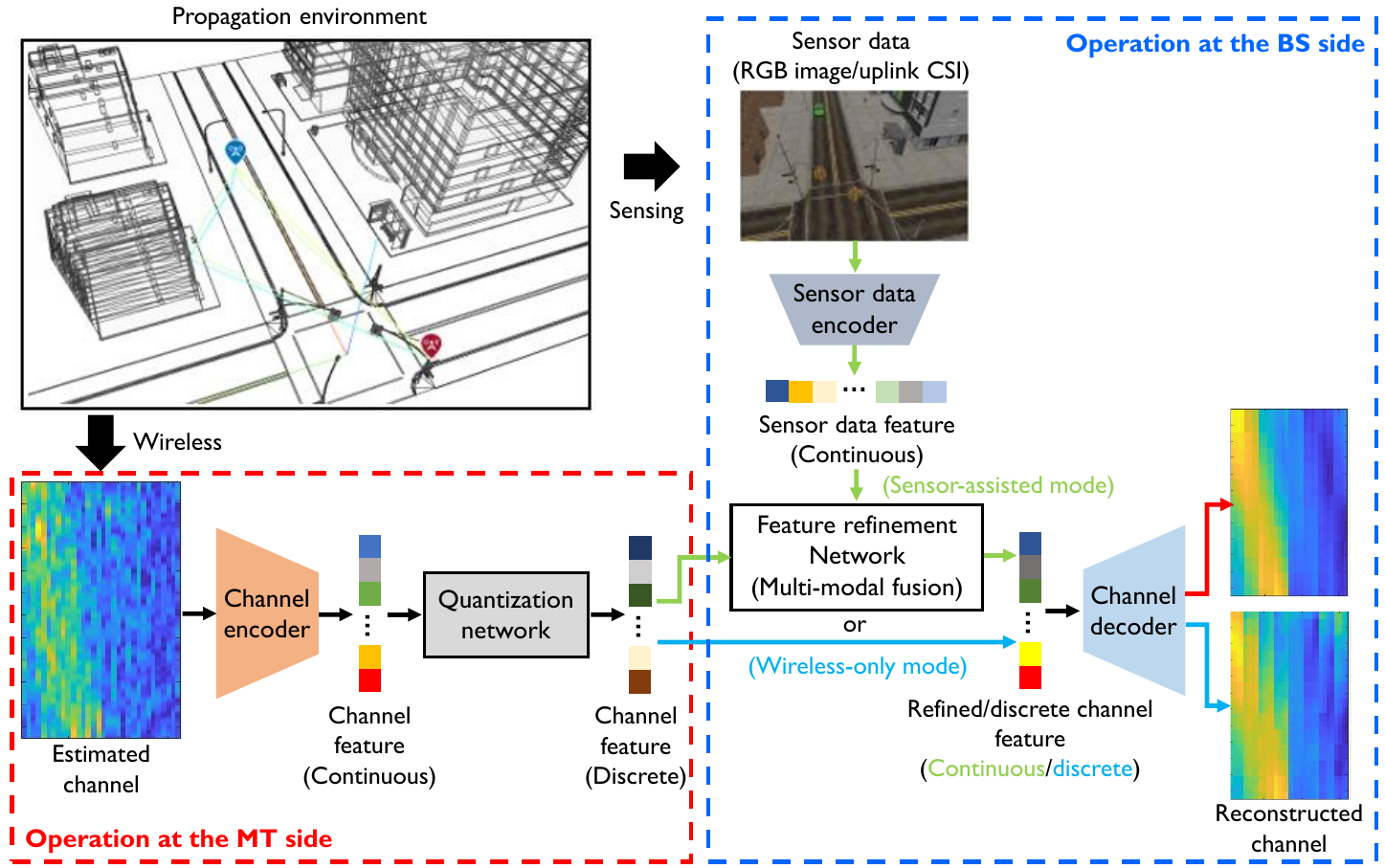}
  \vspace{-0.2cm}
  \caption{Illustration of multi-modal channel reconstruction processes at the BS and the MT.} \label{Fig1}
\end{figure*}

\subsection{Basic Autoencoder Architecture}
We propose an autoencoder comprising an encoder $f_{\text E}(\cdot,B): {\bf H}\!\in\!{\mathbb C}^{N_{\rm t}\times N_{\rm s}}\rightarrow {\bf s}\!\in\!\{0,1\}^B$ and a decoder $f_{\text D}(\cdot): {\bf s}\!\in\!\{0,1\}^B \rightarrow \hat{\bf H}\in\mathbb{C}^{N_{\rm t}\times N_{\rm s}}$ that learns variable-rate discrete representations of the channel. The autoencoder is tailored to leverage a feature vector space amenable to multi-modal fusion. Our design trains channel compression, quantization, and reconstruction collaboratively, distinguishing it from conventional autoencoders that rely on fixed quantization rules \cite{Zeng, Guo, Liang2}. Additionally, unlike autoencoders that require supplementary loss terms (e.g., codebook loss and commitment loss) for quantization network training \cite{Jeon1, VQ-VAE}, our autoencoder is optimized solely with the end-to-end loss function (6).

{\bf 1) CSI generation at the encoder: }The encoder produces a CSI bit stream through feature extraction $\mathcal{E}$, quantization $\mathcal{Q}$, and bit mapping $\mathcal{M}$, i.e., ${\bf s}=f_{\text E}({\bf H},B)=\mathcal{M}(\mathcal{Q}(\mathcal{E}({\bf H})),B))$. The feature extraction process compresses the channel by extracting $N$ real-valued features, i.e., $\mathcal{E}(\cdot): {\bf H}\in\mathbb{C}^{N_{\rm t}\times N_{\rm s}} \rightarrow {\bf z}\in\mathbb{R}^N$, reducing the representing dimension by a ratio of $N/(2N_{\rm t}N_{\rm s})$. The quantization step discretizes each feature with $B_\text{max}\in\mathbb{N}$ bits, generating a sequence of $2^{B_\text{max}}$-ary vectors, i.e., $Q(\cdot):{\bf z}\in\mathbb{R}^N\rightarrow \{{\bf \ell}_1,{\bf \ell}_2,\ldots,{\bf \ell}_N\}$. Then, bit mapping step packs these quantized vectors into a binary stream of length $B$, i.e., $\mathcal{M}(\cdot,B): \{{\bf \ell}_1,{\bf \ell}_2,\ldots,{\bf \ell}_N\}\rightarrow {\bf s}\in\{0,1\}^{B}$.

\begin{figure}[t]
	\centering
  \includegraphics[width=0.6\textwidth]{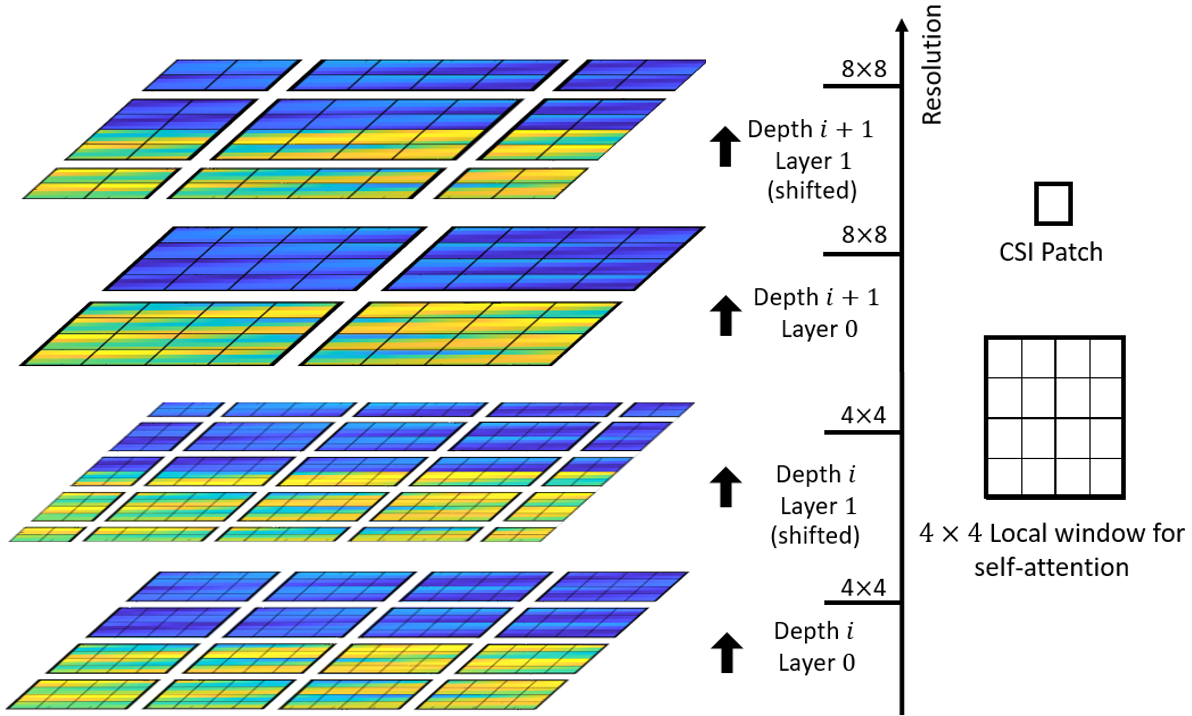}
  \caption{Hierarchical self-attention computation using local windows/shifted-windows.} \label{Fig3}
\end{figure}

Feature extraction forms the backbone of DL-based CSI generation but is highly resource-intensive, posing non-trivial challenges for MTs with constrained memory and processing power. Convolutional neural networks (CNNs) have been popular for their low complexity and effectiveness in capturing local channel characteristics \cite{Liu,Zeng,Guo,Liang2}. However, they struggle with modeling long-range dependencies among spatially distant channel elements. In contrast, Transformer architectures excel at capturing long-range dependencies, but their self-attention mechanism comes with a quadratically increasing computational cost with respect to the channel size \cite{Ju1}. To strike a balance, we employ the Swin Transformer \cite{Swin}, which reduces computational complexity by applying window-based multi-head self-attention (W-MSA) within non-overlapping local windows. For a channel patch resolution of $h\times w$ and an embedding dimension $C$, using $M \times M$ windows yields a complexity of
\begin{align}
    \Omega(\text{W-MSA}) = 4hwC^2+2M^2hwC,
\end{align}
which grows linearly with the channel size. Using non-overlapping windows, W-MSA can only capture the dependency within each window. To provide long-range connections across multiple windows, the subsequent layer computes $\lfloor \frac{M}{2} \rfloor \times \lfloor \frac{M}{2} \rfloor$-shifted window-based MSA. Then, neighboring $2\times2$ patches are concatenated for down sampling (see Fig. 3). This encoding hierarchy enables to efficiently capture long-range dependencies despite using a few network parameters. To align with the 5G NR frame structure \cite{3GPP2}, we set the CSI patch embedding size to $2\times12$, which captures correlations between two adjacent antennas and within one RB. Accordingly, the patch resolution becomes $h\times w =N_{\rm t}/2 \times N_\text{RB}$. We choose $M=4$ to keep the complexity of W-MSA computation low. The network size can be further tuned using the embedding dimensions $[N_{\rm L0}, N_{\rm L1},N_{\rm L2},N_{\rm L3},N_{\rm p}]$ (see Fig. 4). With our design, the feature extraction network produces $N=\frac{N_{\rm t} N_\text{RB}}{128}N_{\rm p}$ continuous-valued channel features.

\begin{figure}[t]
	\centering
  \includegraphics[width=0.6\textwidth]{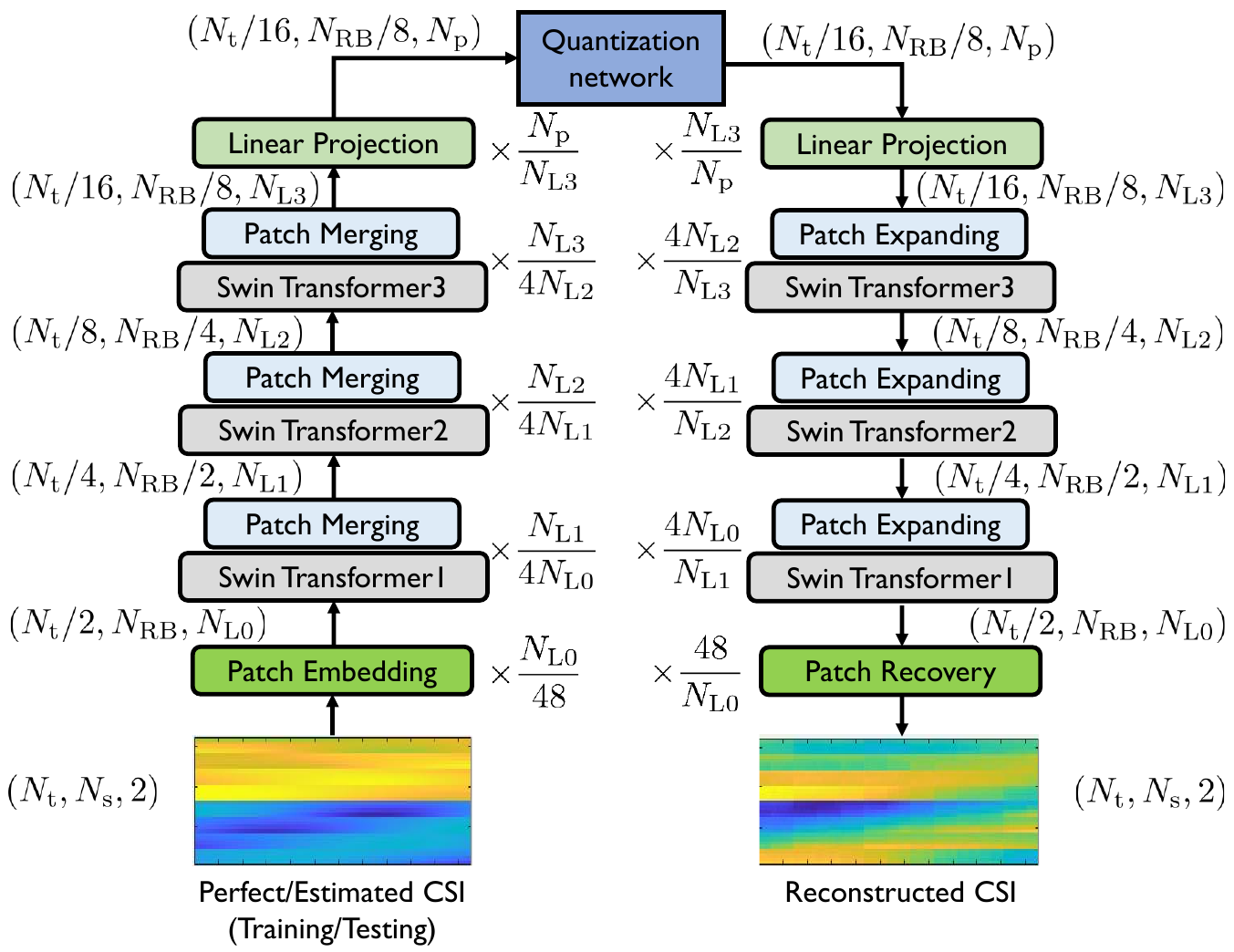}
  \vspace{-0.2cm}
  \caption{Network layers of the autoencoder.}
  \label{Fig3}
\end{figure}

 After feature extraction, a quantization network maps each feature to one of $2^{B_\text{max}}$-ary vectors, i.e., $Q_i(\cdot):z_i\in\mathbb{R}\rightarrow \ {\bf \ell}_i\in\mathcal{L}\subset\{-1,+1\}^{2^{B_\text{max}}-1}$. The quantizer for the $i$th feature $z_i$ is characterized by $2^{B_\text{max}}-1$ trainable quantization boundaries, i.e., $\mathcal{B}_i=\left\{b_i^{(1)}<b_i^{(2)}<,\ldots,<b_i^{(2^{B_\text{max}}-1)}\right\}$. Using $\mathcal{B}_i$, $z_i$ is mapped into ${\bf \ell}_i$ as
\begin{align}
    {\bf \ell}_i&=Q_i(z_i)=\left[\begin{array}{c}\! {\sf sign}(z_i\!-\!b_i^{(1)})\!\\\! {\sf sign}(z_i\!-\!b_i^{(2)})\!\\\! \vdots\!\\\!{\sf sign}(z_i\!-\!b_i^{(2^{B_\text{max}}-1)})\! \end{array}\! \right]
    .\label{quantizer}
\end{align}
A primary hurdle in designing trainable quantization networks is the non-differentiable, zero-gradient nature of the ${\sf sign}(\cdot)$ function, which obstructs gradient flow from the reconstruction loss (6) and prevents direct learning of the embedding rule ${\sf em}(\cdot): {\bf z} \rightarrow {\bf z}_\text{q}$. Conventional autoencoders resolve this problem by introducing auxiliary losses (e.g., $\|{\sf sg}({\bf z})-{\bf z}_\text{q}\|_2$ and $\|{\bf z}-{\sf sg}({\bf z}_\text{q})\|_2$) to reduce quantization error \cite{Jeon1, VQ-VAE}. However, minimizing these embedding losses does not necessarily aid in minimizing the original reconstruction loss. To overcome this limitation, we adopt a surrogate gradient method that approximates ${\sf sign}(\cdot)$ using ${\sf tanh}(\cdot)$ according to
\begin{align}
   \hat{\ell}_{i,k} =
\begin{cases} 
    {\sf tanh}(z_i \!-\! b_i^{(k)}) + {\sf sg}\!\left(\ell_{i,k} \!-\! {\sf tanh}(z_i \!-\! b_i^{(k)})\!\right), & \text{if } \ \ell_{i,k-1}\ell_{i,k}\!=\!-1 \  \text{or } \ \ell_{i,k}\ell_{i,k+1}\!=\!-1, \\ 
    \ell_{i,k}, & \text{otherwise.}
\end{cases}
\end{align}
This formulation selectively propagates gradients through only the critical quantization boundaries $b_i^{(j-1)}$ and/or $b_i^{(j)}$ for which $b_i^{(j-1)}<z_i<b_i^{(j)}$, while masking gradients through non-critical boundaries $b_i^{(k)}$ where $k\!<\!j\!-\!1$ or $k\!>\!j$. The rationale for this design is that adjusting non-critical boundaries does not alter the quantized output signs ${\ell}_{i,k}$ and thus does not affect the training loss. Here, employing ${\sf tanh}(\cdot)$ ensures small modeling error during forward computation, i.e., ${\sf sign}(x)-{\sf tanh}(x)\approx0$. More importantly, its monotonic derivative enable adaptive boundary updates based on their distance to the feature, i.e., $d_i(j)=\|z_i-b_i^{(j)}\|_2$. Specifically, given $b_i^{(j-1)}<z_i<b_i^{(j)}$ and $\frac{\partial L}{\partial \ell_{i,j}}=a$, the surrogate gradient through $b_i^{(j)}$ is
\begin{align}
\frac{\partial L}{ \partial b_i^{(j)}} = \frac{\partial L}{\partial \hat{\ell}_{i,j}} \frac{\partial \hat{\ell}_{i,j}}{ \partial b_i^{(j)}} \approx -a (1-{\sf tanh}^2(z_i-b_i^{(j)})),
\end{align}
leading to aggressive boundary updates when $d_i(j)$ is small. Thanks to the differentiable surrogate function, the network can be trained end-to-end (encoder-quantizer-decoder) exclusively from the desired loss (6). To maintain the order of quantization boundaries ($b_i^{(m)}>b_i^{(n)}$ for $m>n$) during training, each $b_i^{(k)}$ is computed via a cumulative sum of rectified linear outputs:
\begin{align}
    b_i^{(k)}=b_i^{(1)}+\sum_{\ell=2}^{k} {\max}(0, {\beta}_i^{(\ell)}), \ \text{for} \ k\geq 2,
\end{align}
where $b_i^{(1)}$ is the first quantization boundary and ${{\beta}_i^{(\ell)}}\in\mathbb{R}$ for $\ell\in\{2,\ldots,2^{B_\text{max}}-1\}$ are pseudo-quantization intervals. Hence, the quantization layer introduces $N(2^{B_\text{max}}-1)$ trainable parameters overall.

The quantization network $\mathcal{Q}(\cdot)$ produces a set of $2^{B_\text{max}}$-ary vectors, i.e., $\{{\bf \ell}_1,{\bf \ell}_2,\ldots,{\bf \ell}_N\}$, conveying $B=NB_\text{max}$ bits information about ${\bf H}$. The bit mapper $\mathcal{M}(\cdot,NB_\text{max})$ converts these vectors into a bit stream ${\bf s}=[{\bf s}_1^{\top}, {\bf s}_2^{\top},\ldots,{\bf s}_N^{\top}]^{\top}$ through
\begin{align}
    {\bf s}_i=\text{D2B}_{B_{\text{max}}}\Big(\sum_{j=1}^{2^{B_\text{max}}-1} {\sf 1}_{{\bf \ell}_{i,j}=1} \Big),
\end{align}
where $\text{D2B}_{B_{\text{max}}}(\cdot)$ denotes a $B_{\text{max}}$ bit decimal-to-binary converter. In Subsection B, we extend this one-to-one mapping to accommodate a variable target rate $B<NB_\text{max}$, i.e., $\mathcal{M}(\cdot,B): \{{\bf \ell}_1,{\bf \ell}_2,\ldots,{\bf \ell}_N\}\rightarrow {\bf s}\in\{0,1\}^B$.

{\bf 2) CSI reconstruction at the decoder: }The decoder at the BS reconstructs the channel through bit demapping $\mathcal{M}^{-1}$, feature space demapping $\mathcal{D}$, and channel recovery $\mathcal{C}$, i.e., $\hat{\bf H}=f_{\text D}({\bf s}) = \mathcal{C}(\mathcal{D}(\mathcal{M}^{-1}({\bf s})))$. The bit demapper $\mathcal{M}^{-1}(\cdot): {\bf s}\in\{0,1\}^B\rightarrow \{{\bf \ell}_1,{\bf \ell}_2,\ldots,{\bf \ell}_N\}$ performs the inverse operation of $\mathcal{M}(\cdot,B)$, based on the size of ${\bf s}$. The feature space demapper $\mathcal{D}(\cdot):\{{\bf \ell}_1,{\bf \ell}_2,\ldots,{\bf \ell}_N\}\rightarrow {\bf z}({\bf s})\in\{1-2^{B_\text{max}},2-2^{B_\text{max}}\ldots,2^{B_\text{max}}-1\}^N$ computes the discrete representation counterpart (quantization level vector) of ${\bf z}$ by
\begin{align}
    {\bf z}({\bf s})=[{\bf 1}^{\top}{\bf \ell}_1, {\bf 1}^{\top}{\bf \ell}_2, \ldots, {\bf 1}^{\top}{\bf \ell}_N]^{\top}.
\end{align}
By taking the summation over the comparator output, i.e., $z({\bf s})_i={\bf 1}^{\top}{\bf \ell}_i$, quantization level outputs can implicitly carry magnitude information of unquantized inputs. Using ${\bf z}({\bf s})$, the channel is reconstructed through a neural network $\mathcal{C}(\cdot): {\bf z}({\bf s})\rightarrow \hat{\bf H}\in\mathbb{C}^{N_{\rm t}\times N_{\rm s}}$, which has symmetrical network layers to $\mathcal{E}$ (see Fig. 4).

\subsection{Variable-Rate Quantization}
In practical systems, CSI feedback length $B$ is dictated by radio resource control parameters \cite{3GPP1}. However, most DL-based CSI feedback schemes are unable to flexibly generate CSI at variable rates, because the feature extractor $\mathcal{E}$ always produces a fixed-dimensional vector ${\bf z}\in\mathbb{R}^N$. To enable variable-rate CSI generation, we introduce a downsampling rule $\mathcal{M}(\cdot,B): \{{\bf \ell}_1,{\bf \ell}_2,\ldots,{\bf \ell}_N\}\rightarrow {\bf s}\in\{0,1\}^B$ and train the quantizer $\mathcal{Q}$ to comply with this rule.

Key idea to flexibly generate $B<NB_\text{max}$ bit CSI is to uniformly downsample the quantization boundaries in $\mathcal{B}_i$. Suppose the feedback length is $B=\sum_{i=1}^{N}B_i$, and the $i$th feature $z_i$ is assigned $B_i<B_{\text{max}}$ bits. We define a downsampling factor corresponding to the $i$th feature by $\alpha_i = 2^{B_{\text{max}}-B_i}$. Then, we allocate $B_i$ bits to $z_i$ by setting the effective quantization boundaries $\mathcal{B}_{i,\text{eff}}\subset \mathcal{B}_{i}$ as
\begin{align}
    \mathcal{B}_{i,\text{eff}}=\left\{b_i^{(\alpha_i)}, b_i^{(2\alpha_i)}\ldots,b_i^{((2^{B_i}-1)\alpha_i)}\right\}.
\end{align}
The remaining quantization boundaries in $\mathcal{B}_{i}\setminus\mathcal{B}_{i,\text{eff}}$ are deactivated, causing partial information erasure in ${\bf \ell}_i$. Nonetheless, most of the entries in ${\bf \ell}_i$ can be perfectly restored using the known values at the selected boundaries. Specifically, since $\mathcal{B}_{i}$ is an ordered set, ${\bf \ell}_{i,{j_1}}=1$ guarantees ${\bf \ell}_{i,{j_2}}=1$ for $j_1>j_2$. Similarly, ${\bf \ell}_{i,{j_1}}=-1$ guarantees ${\bf \ell}_{i,{j_2}}=-1$ for $j_1<j_2$. Accordingly, all erasures, except for the $\alpha_i-1$ entries among ${\bf \ell}_i$ can be inferred without error. For the remaining $\alpha_i-1$ erasures, we simply interpolate the values by ${\bf \ell}_{i,{j}}=0$. As a result, this downsample-then-interpolate strategy allows the $B_i$ bit quantizer to emulate the behavior of $B_{\text{max}}$ bit quantizer. The corresponding bit stream for the $i$th feature is
\begin{align}
    {\bf s}_i=\text{D2B}_{B_i}\Big(\sum_{j=1}^{2^{B_i}-1} {\sf 1}_{{\bf \ell}_{i,\alpha_i j}=1} \Big).
\end{align}

{\bf Example: }Suppose $B_{\text{max}}=3$. In this case, the quantizer for the $i$th feature $z_i$ is characterized by $\mathcal{B}_i=\{b_i^{(1)}\!<\!b_i^{(2)}\!<,\ldots,\!<\!b_i^{(7)}\}$. For reduced rates, we get $\mathcal{B}_{i,2\text{bits}}=\{b_i^{(2)}\!<\!b_i^{(4)}\!<\!b_i^{(6)}\}$, and $\mathcal{B}_{i,1\text{bit}}=\{b_i^{(4)}\}$. When $b_i^{(2)}<z_i<b_i^{(3)}$, we obtain
\begin{align}
    {\bf \ell}_i &= [+1 \ +\!1 \ -\!1 \ -\!1 \  -\!1  \ -\!1  \ -\!1]^{\top} \ (\text{3bits}) \nonumber\\
    {\bf \ell}_i &= [+1 \ +\!1 \ \ \ 0 \ -\!1 \  -\!1  \ -\!1  \ -\!1]^{\top} \ (\text{2bits}) \nonumber \\
{\bf \ell}_i &= [\ 0 \ \ \ \ \!0 \ \ \ \ \! 0 \ -\!1 \  -\!1  \ -\!1  \ -\!1]^{\top} \ (\text{1bit}).
\end{align}
The bit streams corresponding to the $i$th feature are
\begin{align}
    {\bf s}_i &= [0 \ 1 \ 0]^{\top} \ (\text{3bits}), \nonumber\\
    {\bf s}_i &= [0 \ 1]^{\top} \ (\text{2bits}), \nonumber\\
    {\bf s}_i &= [0]^{\top} \ (\text{1bit}).
\end{align}

The proposed downsampling rule allows generating CSI with any desired length $B\in[N, NB_{\text{max}}]$. But, the reconstruction loss (6) optimizes the network only for a single feedback length. To accommodate multiple feedback lengths, we train the encoder–decoder pair using a weighted-average loss:
\begin{align}
    L_{\text{WA}}\left({\bf H}, \{\hat{\bf H}^{(i)}\}_{i=1}^{K}, \{B^{(i)}\}_{i=1}^{K}, \gamma\right)=\frac{1}{\sum_{i=1}^{K}\gamma^{B^{(i)}}}\sum_{i=1}^{K} {\gamma}^{B^{(i)}}L\left({\bf H}, \hat{\bf H}^{(i)}\right), \label{Weightedloss}
\end{align}
where $\hat{\bf H}^{(i)}$ is the reconstructed channel using $B^{(i)}$ bits and $\gamma>1$ is a hyper-parameter that assigns larger weight for higher quantization rate. With this weighted-average loss, the quantization boundaries in $\mathcal{Q}$ are optimized across all target feedback rates.

{\bf Remark 1 (Disjoint quantization levels at variable-rates): }The proposed quantization network focuses on classifying the range of continuous-valued inputs. Using $B_i$ bits, the set of feasible classes is
\begin{align}
z({\bf s})_i\in\mathcal{S}(B_i)=\left\{2\alpha_i \left\lfloor\frac{1}{\alpha_i} \left(n - \frac{2^{B_{\text{max}}}-1}{2}\right)\right\rfloor+\alpha_i\right\}_{n=0}^{2^{B_{\text{max}}}-1}.
\end{align}
A key feature of the downsample-then-interpolate strategy is the formation of disjoint sets of classes for different resolutions, i.e., $\mathcal{S}(B_i) \cap\mathcal{S}(B_j)=\phi$ for $B_i\neq B_j$. In contrast, in quantization networks that explicitly approximate the continuous-valued inputs \cite{Jeon1}, a quantized output chosen from a low rate codebook is also a codeword in a higher rate codebook, i.e., ${\bf z}_\text{q}\in\mathcal{C}_{\text {low}}\subset \mathcal{C}_{\text {high}}$. The disjoint output spaces at different rates are advantageous at the decoder, particularly during multi-modal channel reconstruction. Specifically, since the channel feature vector ${\bf z}({\bf s})$ informs the reliability of CSI feedback, the multi-modal fusion network can autonomously and adaptively evaluate the significance of sensor data across various rates.

{\bf Remark 2 (Efficiency of the proposed quantizer): }The proposed quantizer provides several notable advantages over VQ-based networks. First, it requires only $N(2^{B_\text{max}}-1)$ parameters for variable-rate quantization, which amounts to no more than a few hundred. Second, the quantization mapping is computationally efficient, implemented with a simple comparator as described in \eqref{quantizer}. In contrast, when the embedding size is $N_{\text{size}}$, VQ requires $N2^{N_{\text{size}}B_{\text {max}}}$ parameters $(N_{\text{size}}\!\!\gg\!\! 1)$ to describe a quantization codebook. This can easily result in parameter counts exceeding several hundred thousand. Additionally, the quantization mapping in VQ methods involves an exhaustive search through a codebook of size $N_{\text{size}}\times2^{N_{\text{size}}B_{\text {max}}}$. Consequently, the proposed element-wise, variable-rate quantization is particularly well-suited for practical MTs with limited memory and power resources.

\subsection{Multi-Modal Fusion}
CSI distortion arising from channel estimation noise, as well as compression and quantization during channel feedback, are critical factors that degrade downlink beamforming gains. To mitigate this, we propose restoring the CSI by leveraging auxiliary data available at the BS (e.g., RGB images and uplink CSI).

{\bf 1) Channel-relevant feature extraction from sensor data: }Given sensor data ${\bf D}$, the BS encodes this data to extract channel-relevant features, i.e., $g_{\text E}(\cdot): {\bf D}\rightarrow {\bf z}({\bf D})\in{\mathbb R}^M$. These features contain positional information of the MTs and potential channel clusters, making them valuable for channel reconstruction. Traditional computer vision-only beamforming techniques rely on supervised learning for feature extraction, requiring pixel-level MT location labels \cite{Kim, Ahn2}. They identify MT positions through object detection or semantic segmentation, then convert these into LOS ray parameters (e.g., distances, angles). Under ideal conditions (MT with an isotropic antenna, stationary MT, and no NLOS path), the translated channel parameters can be mapped to an optimal beamforming matrix. However, generalizing supervised learning-based approach to multi-path channels is challenging, as it requires labeled data for all channel reflectors. Moreover, the ray parameters may be redundant when CSI feedback is already available. To address these issues, we propose a self-supervised learning approach that learns non-redundant features directly from ${\bf D}$ given the CSI. Fig. 5 depicts the image-based feature extraction network, whose output dimension $M=\frac{N_{\text H}N_{\text W}}{64n_{\text H}n_{\text W}}N_{\text e}$ can be controlled by adjusting the dimensions $[N_{\ell1},N_{\ell2},N_{\ell3},N_{\text e}]$. Here, the final embedding dimension $N_{\text e}$ is chosen among $\{N_{\rm p}, N_{\rm L3},N_{\rm L2},N_{\rm L1}\}$ for subsequent multi-modal fusion.

\begin{figure}[t]
	\centering
  \includegraphics[width=0.55\textwidth]{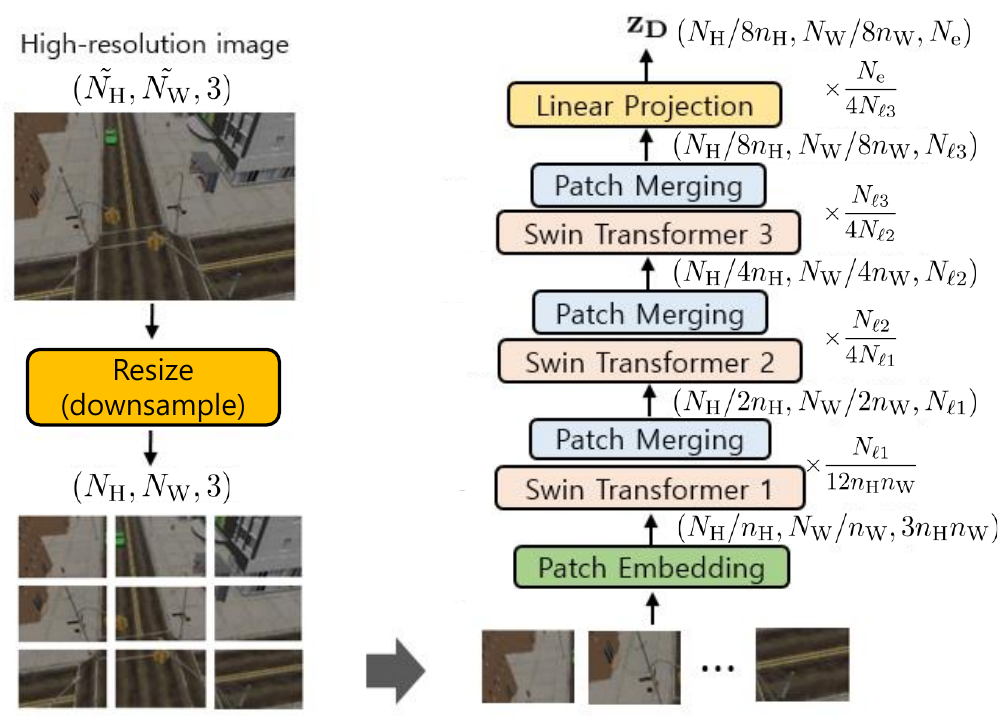}
  \caption{Architecture of the channel-relevant feature extraction network.} \label{Fig4}
\end{figure}

{\bf 2) Fusion of CSI and sensor data features: }Upon receiving CSI feedback and encoding sensor data, the BS has access to an $N$-dimensional channel feature ${\bf z}({\bf s})$ and an $M$-dimensional sensor data feature ${\bf z}({\bf D})$. However, the channel recovery network $\mathcal{C}$ requires an $N$-dimensional vector as input. To reconcile this dimensional mismatch while leveraging both data sources, the BS performs feature vector refinement, i.e., ${\bf z}_{\text r}=\mathcal{R}\left({\bf z}({\bf s}), {\bf z}({\bf D})\right)\in\mathbb{R}^N$. This refinement network restores the lossy, discrete-valued channel feature ${\bf z}({\bf s})$ into a continuous-valued channel feature ${\bf z}_{\text r}$ by incorporating complementary information extracted from ${\bf D}$. For this task, we employ a multi-modal Transformer based on early concatenation \cite{MM}, which enables the model to learn all pairwise interactions between channel and sensor data features.

\begin{figure}[t]
	\centering
  \includegraphics[width=0.62\textwidth]{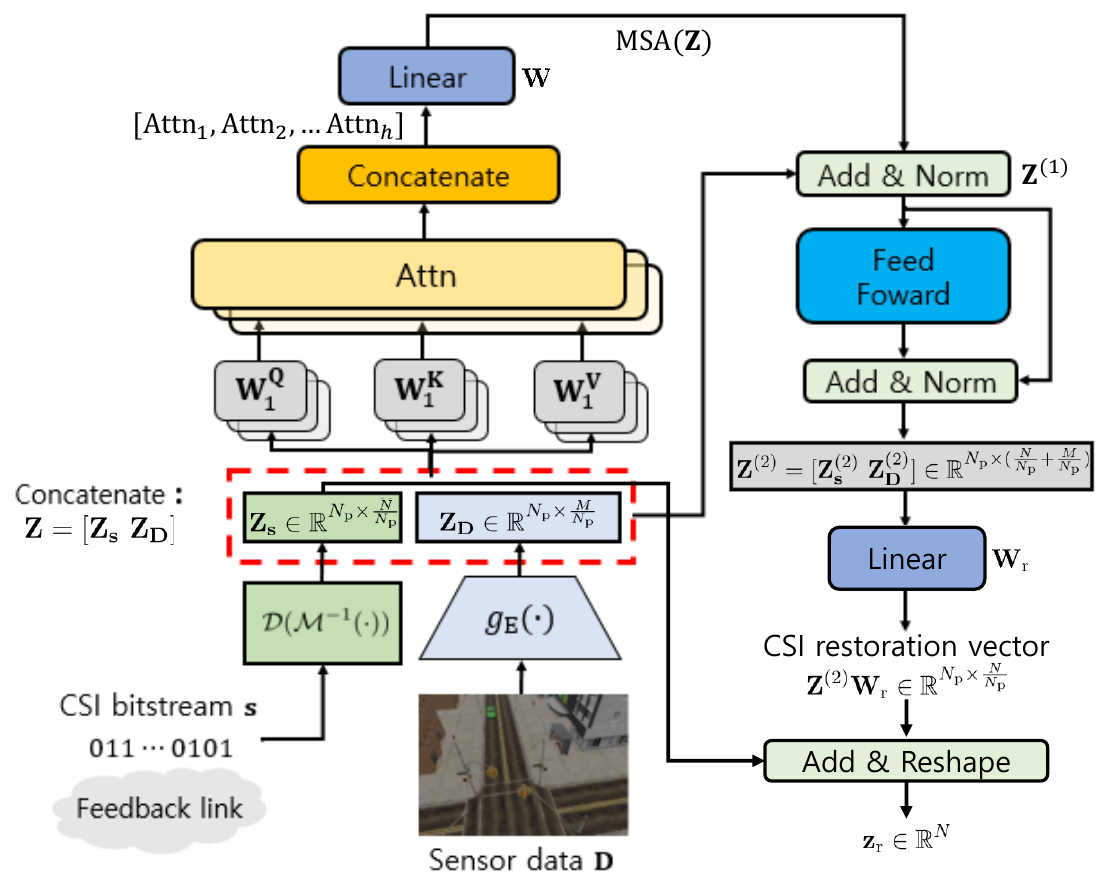}
  \caption{Sensor data-assisted channel feature vector refinement procedure using multi-modal Transformer.} \label{Fig4}
\end{figure}

We describe sensor data-assisted CSI restoration process that is performed after the quantization process ($N_{\text e}=N_{\rm p}$). The refinement network $\mathcal{R}$ reshapes ${\bf z}({\bf s})$ and ${\bf z}({\bf D})$ into ${\bf Z}_{\bf s}\in\mathbb{R}^{N_{\rm p} \times \frac{N}{N_{\rm p}}}$ and ${\bf Z}_{\bf D}\in\mathbb{R}^{N_{\rm p} \times \frac{M}{N_{\rm p}}}$, then concatenates them into ${\bf Z}_{\text{cat}}=\left[{\bf Z}_{\bf s} \ {\bf Z}_{\bf D}\right]\in\mathbb{R}^{N_{\rm p} \times {(\frac{N}{N_{\rm p}}+\frac{M}{N_{\rm p}})}}$. To incorporate positional information, a learnable bias ${\bf B}_\text{pos}\in\mathbb{R}^{N_{\rm p} \times {(\frac{N}{N_{\rm p}}+\frac{M}{N_{\rm p}})}}$ is added to the concatenated features, i.e., ${\bf Z}={\bf Z}_{\text{cat}}+{\bf B}_\text{pos}$. To capture cross-modal dependencies between CSI and sensor data, ${\bf Z}$ is embedded into multiple queries, keys, and values, as follows:
\begin{align}
    {\bf Q}_i={\bf W}_i^{\bf Q}{\bf Z}, \  {\bf K}_i={\bf W}_i^{\bf K}{\bf Z}, \ {\bf V}_i={\bf W}_i^{\bf V}{\bf Z} \ \ \text{for} \ \ i\in\{1,2,\ldots,h\}, \label{qkvmap}
\end{align}
where ${\bf W}_i^{\bf Q}, {\bf W}_i^{\bf K}, {\bf W}_i^{\bf V} \in \mathbb{R}^{d_\text{dim}\times N_{\rm p}}$ and $N_{\rm p}=hd_\text{dim}$. Using \eqref{qkvmap}, the attention for the $i$th head is computed as
\begin{align}
    {\text{Attn}}_i = {\sf{softmax}} (\frac{{\bf Q}_{i}^{\top}{\bf K}_{i}}{\sqrt{d_{\text{dim}}}}) {\bf V}_i^{\top}\in \mathbb{R}^{(\frac{N}{N_{\rm p}}+\frac{M}{N_{\rm p}}) \times d_{\text{dim}}}.
\end{align}
This attention score quantifies the importance of each feature in relation to the others. After calculating the attention scores, the outputs from the multiple heads are concatenated and linearly transformed:
\begin{align}
{\text {MSA}}({\bf Z})={\bf W}\left[{\text {Attn}}_1,{\text {Attn}}_2,\ldots,{\text {Attn}}_h\right]^{\top}\in\mathbb{R}^{N_{\rm p} \times {(\frac{N}{N_{\rm p}}+\frac{M}{N_{\rm p}})}},
\end{align}
where ${\bf W}\in\mathbb{R}^{N_{\rm p}\times N_{\rm p}}$. The model then applies residual connections, layer normalization, and a feed-forward network:
\begin{align}
    &{\bf Z}^{(1)}={\text{LN}}({\bf Z}+{\text {MSA}}({\bf Z})), \nonumber\\
    &{\bf Z}^{(2)} ={\text{LN}}( {\text{MLP}}({\bf Z}^{(1)})+{\bf Z}^{(1)}).
\end{align}
Using the jointly encoded features ${\bf Z}^{(2)}\in\mathbb{R}^{N_{\rm p} \times {(\frac{N}{N_{\rm p}}+\frac{M}{N_{\rm p}})}}$, we compensate for CSI distortion as
\begin{align}
    {\bf Z}_{\text r}={\bf Z}_{\bf s}+{\bf Z}^{(2)}{\bf W}_{\text r}\in\mathbb{R}^{N_{\rm p} \times \frac{N}{N_{\rm p}}},
\end{align}
where ${\bf W}_{\text r}\in\mathbb{R}^{{(\frac{N}{N_{\rm p}}+\frac{M}{N_{\rm p}})}\times \frac{N}{N_{\rm p}}}$. Finally, we reshape ${\bf Z}_{\text r}$ into ${\bf z}_\text{r}\in\mathbb{R}^N$ and feed it into $\mathcal{C}$. The feature vector refinement process is illustrated in Fig. 6.

{\bf Remark 3 (Multi-modal fusion using higher embedding dimensions): }Equations (22)–(26) describe the CSI restoration process at the autoencoder’s bottleneck, which utilizes an embedding dimension of $N_{\text e}=N_{\rm p}$. In principle, multi-modal fusion can also be performed deeper in the decoding pipeline, within $\mathcal{C}(\cdot)$. For instance, fusion operations could occur after the linear projection layer in the decoder, where the embedding dimension is $N_{\text e}=N_{\rm L3}$. Leveraging higher embedding dimension allows sensor data to provide richer environment-aware information about the channel, leading to better channel reconstruction accuracy. However, this approach comes at the cost of increased network complexity.

\section{Dataset Generation and Network Training}
In this section, we describe sensor data-wireless channel paired dataset generation process and two-stage network training method for the multi-modal, variable-rate CSI reconstruction framework.

\begin{figure*}[t]
	\centering
  \includegraphics[width=1\textwidth]{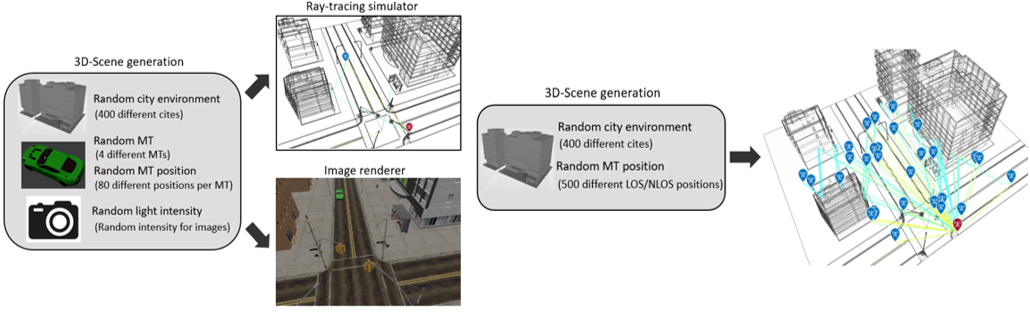}
  \caption{RGB image-wireless channel paired dataset generation for LOS environments and wireless channel-only dataset generation for LOS/NLOS environments.} 
  \label{Fig3}
\end{figure*}

\subsection{Sensor Data-Wireless Channel Paired Dataset}
The foundation of our multi-modal CSI reconstruction framework is a large, high quality dataset pairing sensor data with wireless channel measurements. Due to the data-driven nature of DL, channel restoration accuracy can depend heavily on the quality of a training dataset. However, acquiring a large real-world channel measurement data is extremely time-consuming and expensive. To avoid this difficulty, we simulate real-world sensor-assisted wireless scenarios using a synthetic dataset. Our synthetic dataset generation is scalable, parametric, and noise-free. We generate our own synthetic dataset from the following procedures:

\begin{itemize}
    \item {\bf Scenario modeling and image rendering: } We create virtual 3D city environments using the 3D modeling software, Blender. Each scene includes various objects such as buildings, cars, roads, traffic lights, bus stations, trees and street lamps. The scene can be parameterized using $(P_1, P_2, C, L, S)$, where $P_1$ and $P_2$ represents the parts of the city drawn from $20$ different pre-designed city segments. $C$ denotes the car model, chosen from $4$ different designs. $L=(L_x, L_y, L_z)$ specifies the local coordinates of the car within the scene. $S$ represents the intensity of sun light, modeling the virtual timing of day. Sample images for fixed $(P_1,P_2,C)$ and random $(L,S)$ values are shown in Fig. 8.
    
    \begin{figure*}[t]
	\centering
  \includegraphics[width=0.9\textwidth]{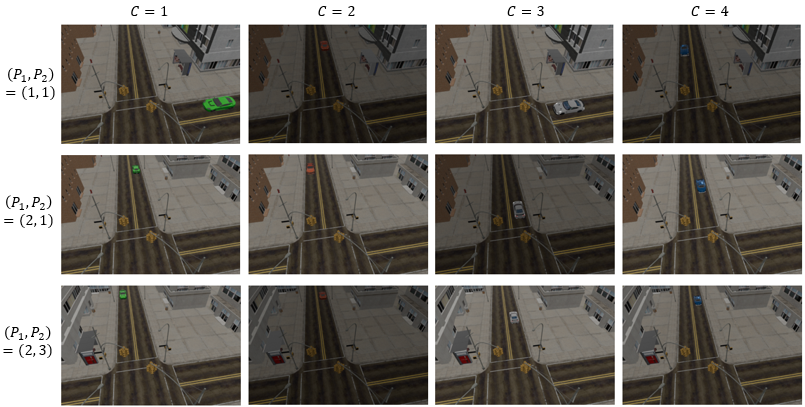}
  \caption{Sample RGB images for fixed $(P_1, P_2, C)$ and random $(L,S)$ values.} \label{Fig4}
\end{figure*}

\item {\bf Channel generation based on ray-tracing: } In the simulated scene, the car object is assumed to be the MT of interest. For the RGB image-downlink channel paired dataset, the BS and the camera sensor are assume to be co-located at $15$ meters height. The orientation (e.g., azimuth and elevation) of the BS arrays and the camera's field of view are fixed, while the orientation of the MT antenna is chosen randomly. Under these settings, we export the 3D city models and perform ray-tracing simulations. For each scene, multi-path channel parameters including path gains, delays, angle of arrival, and angle of departure are extracted at the carrier frequency $f_c$. We limit the reflection order of rays to 2. Once the channel parameters are obtained, we construct spatial-time domain channel impulse responses (CIRs) according to the clustered delay line channel generation process \cite{3GPP3}. Finally, OFDM is applied to convert the CIR into a spatial-frequency domain channel ${\bf H}\in{\mathbb C}^{N_{\rm t}\times N_{\rm s}}$.
\end{itemize}
By following the two procedures, we generate $128,000$ pairs of RGB images and wireless channels.

\subsection{Two-Stage Training using Heterogeneous Datasets}
One major limitation of our RGB image-downlink channel paired dataset is the positional bias of the MTs toward LOS locations. Specifically, restricting MTs to appear within the camera’s field of view results in predominantly LOS channel conditions. However, the autoencoder should also perform well in non-LOS (NLOS) scenarios where image data may be unavailable at the BS. To ensure unbiased training across both types of channels, we utilize an additional dataset comprising wireless channel-only instances. In this supplementary dataset, MTs are positioned at arbitrary locations with and without LOS paths. At each MT location, we generate a pair of downlink channel and uplink channel that shares the same geometrical ray parameters. Using the two datasets, we train our multi-modal, variable-rate autoencoder network as follows:

\begin{itemize}
    \item {\bf Stage 1: }We start by independently training the CSI bit stream generator $f_{\text E}(\cdot,B)$ and the channel recovery network $\mathcal{C}(\cdot)$ without sensor data. The aim of this stage is to learn a universal DL-assisted CSI codebook that functions effectively in both LOS and NLOS channel conditions. To achieve this, we use downlink channels from the wireless-only dataset. During training, we use perfect CSI as both the input and target of the network. The forward computation and the training loss for Stage 1 are as follows:
    \begin{align}
        &\text{Encoder: } {\bf s}^{(i)}=f_{\text E}({\bf H},B^{(i)}), \nonumber\\
        &\text{Decoder: } \hat{\bf H}^{(i)}=f_{\text D}({\bf s}^{(i)}),  \nonumber\\
        &\text{Loss: } L_{\text{WA}}\left({\bf H}, \{\hat{\bf H}^{(i)}\}_{i=1}^{K}, \{B^{(i)}\}_{i=1}^{K}, \gamma\right).
    \end{align}
    
    \item {\bf Stage 2: } We then conduct sensor data-dependent transfer learning for the channel-relevant feature extraction network $g_{\text E}$ and the feature vector refinement network $\mathcal{R}$. We use the sensor data-wireless channel paired dataset. The parameters trained during Stage 1 remain frozen to preserve the foundational feature representations. The forward computation and the training loss for Stage 2 are as follows:
\begin{align}
        &\text{Encoder: } {\bf s}^{(i)}=f_{\text E}({\bf H},B^{(i)}),  \nonumber\\
        &\text{Refinement: } {\bf z}_{\text r}^{(i)}=\mathcal{R}\left(\mathcal{D}(\mathcal{M}^{-1}({\bf s}^{(i)})),g_{\text E}({\bf D})\right),
        \nonumber\\
&\text{Decoder: } \hat{\bf H}^{(i)} = \mathcal{C}({\bf z}_{\text r}^{(i)}), \nonumber\\
&\text{Loss: } L_{\text{WA}}\left({\bf H}, \{\hat{\bf H}^{(i)}\}_{i=1}^{K}, \{B^{(i)}\}_{i=1}^{K}, \gamma\right).
\end{align}
\end{itemize}
Through this two-stage learning based on heterogeneous datasets, the network acquires both a site-independent, variable-rate CSI feedback codebook and a site-dependent, multi-modal channel reconstruction capability. Consequently, the BS can reconstruct the channel in either a wireless-only mode or a sensor data-assisted mode.

%
%

\section{Simulation Results}
In this section, we evaluate the performance of our multi-modal, variable-rate autoencoder in terms of channel reconstruction loss and beamforming gain. The encoder and decoder employ embedding dimensions $[N_{\rm L0},N_{\rm L1},N_{\rm L2},N_{\rm L3},N_{\rm p}] = [24, 32, 32, 32, 4]$. Under this setup, the output dimension of the bottleneck layer is $N=48$. We set the maximum resolution for element-wise quantization as $B_\text{max}=3$, allowing the generation of CSI with lengths ranging from $48$ to $144$ bits. We list MIMO-OFDM simulation parameters in Table. I.

\begin{table}[h]
\centering
\caption{Simulation parameters for CDL channel model.}
\vspace{-0.3cm}
\begin{tabular}[t]{|c|c|}
\hline
\hline
Parameter & Value\\
\hline
Downlink carrier frequency $(f_{c})$ & 28GHz \\
Uplink carrier frequency $(f_{c})$ & 27GHz \\
Subcarrier spacing (SCS) & 60kHz \\
Number of resource blocks $N_\text{RB}$ & 48 \\
Number of subcarriers $(N_s)$ & 576 \\
Fast Fourier transform size $(N_\text{FFT})$ & 1024 \\
Sampling rate $(1/T_s)$ & 61440000 \\
BS array & $4\times 4$ UPA ($\pm45^{\circ}$ polarization) \\
MT array & Isotropic antenna \\
\hline
\hline
\end{tabular}
\end{table}

\subsection{Wireless-Only Channel Reconstruction}
We consider a scenario, where the BS performs channel reconstruction in a wireless feedback-only mode. To highlight the impact of quantization network design, we compare the following benchmarks, while fixing the encoder and decoder (see Table II):

\begin{itemize}
    \item {\bf Proposed autoencoder: }Channel feature dimension $N=48$ and resolution $B_\text{max}=3$.
    \item {\bf Nested VQ autoencoder ($N_{\text{size}}=4$): }The autoencoder learns CSI codebook embedding rule, i.e., ${\sf em}(\cdot): {\bf z} \rightarrow {\bf z}_\text{q}$ to minimize
   \begin{align}
    L_{\text{VQ}}=\frac{1}{\sum_{i=1}^{K}\gamma^{B^{(i)}}}\sum_{i=1}^{K} {\gamma}^{B^{(i)}} \left( L\left({\bf H}, \hat{\bf H}^{(i)}\right)+\|{\sf sg}({\bf z})-{\bf z}^{(i)}_{\text q}\|_2+\frac{1}{4}\|{\bf z}-{\sf sg}({\bf z}^{(i)}_{\text q})\|_2\right),
\end{align}
where ${\bf z}^{(i)}_{\text q}$ is a quantized output using input ${\bf z}$ at rate $B^{(i)}$. 
\end{itemize}
 \begin{table}[h]
\caption{Autoencoder model description}
\centering
\renewcommand{\arraystretch}{1.1}
\begin{tabular}{|c|c|c|c|}
  \hline
  \multicolumn{1}{|c|}{} & \multicolumn{3}{c|}{Number of parameters}  \\ \cline{2-4}
  \multicolumn{1}{|c|}{} & Encoder & Quantizer & Decoder  \\ \hline
    Proposed & 100,776 & 336 & 100,766 \\ \hline
  Nested VQ ($N_{\text{size}}=4$) & 100,776 & 196,608 & 100,766 \\ \hline
\end{tabular}
\end{table}
For training, we use downlink channel samples in the wireless-only dataset. We separate $200,000$ channel samples into $150,000$ training data, $30,000$ validation data and $20,000$ test data. The batch size is $N_{\text{Batch}}=200$ and the network is trained for $500$ epochs. We use the Adam optimizer with $10^{-3}$ learning rate. For variable-rate network training, we set target feedback lengths as $B=\{48,72,96,120,144\}$. The hyper parameter for variable-rate training is $\gamma=2^{1/96}$. All networks are trained using perfect CSI input. Then, we use imperfect CSI, estimated at SNR =$\{-10,-5,0\}$dB during evaluation. Here, the estimated channels are obtained by using the CSI-reference signal (CSI-RS) \cite{3GPP2}.

\begin{figure}[t]
	\centering
  \includegraphics[width=0.65\textwidth]{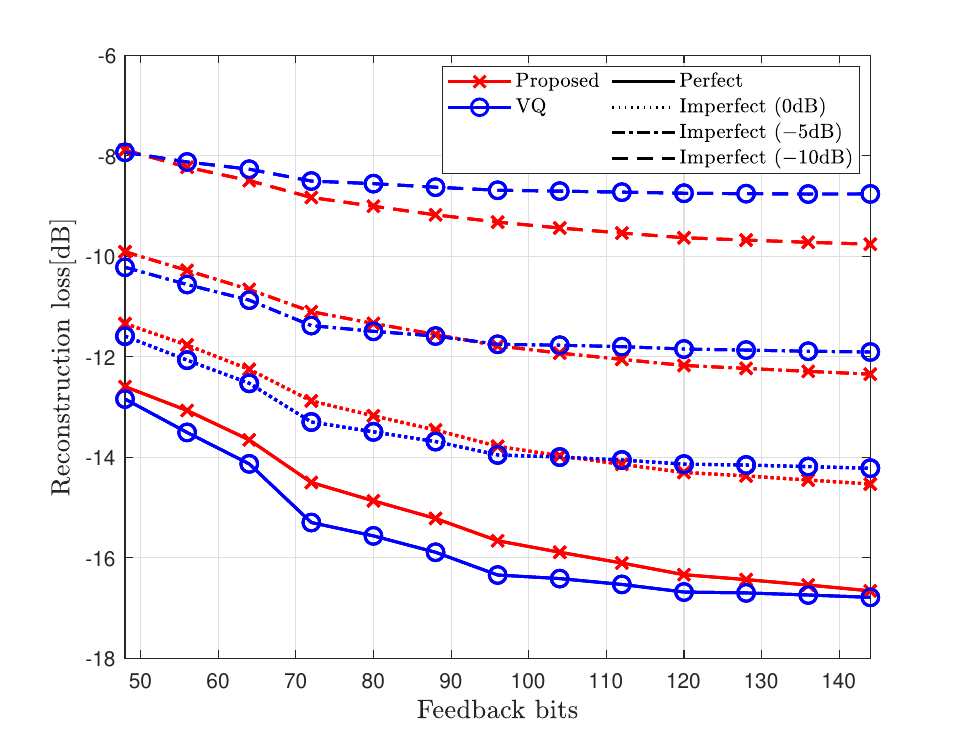}
  \caption{Channel reconstruction losses using benchmarks with respect to various feedback rates at SNR $=\{-10,-5,0\}$dB.}
\end{figure}

Fig. 9 shows the channel reconstruction losses under varying feedback rates and SNRs. In both models, reconstruction accuracy improves with higher feedback rates or SNR levels. A key observation is the superior generalization capability of the proposed quantizer at untrained feedback rates. In contrast to the nested VQ, whose reconstruction accuracy abruptly declines at untrained rates ($B=\{56,64,80,88,104,112,128,136\}$), the proposed network achieves a more consistent performance transition. This ability to adapt in real time is especially appealing, considering that training quantization networks over multiple rates linearly increases the training overhead. Another notable observation is the robustness of the proposed quantizer against channel estimation errors. Despite utilizing significantly fewer network parameters, the proposed model achieves better reconstruction accuracy at low SNR and high feedback rates. This is because implicitly characterizing the range of inputs using (15) is more resilient to noise than explicitly approximating the continuous-valued inputs.


Fig. 10 compares the cumulative distribution functions (CDFs) of precoded channel gains using the following benchmarks:
\begin{itemize}
\item {\bf Ideal beamforming: }Beamforming vector for the $n$th subcarrier is ${\bf p}_n={{\bf h}_n}/{\|{\bf h}_n\|_2}$.

\item {\bf Proposed: }Beamforming vector for the $n$th subcarrier is ${\bf p}_n={\hat{\bf h}_n}/{\|\hat{\bf h}_n\|_2}$.

\item {\bf Nested VQ: }Beamforming vector for the $n$th subcarrier is ${\bf p}_n={\hat{\bf h}_n}/{\|\hat{\bf h}_n\|_2}$.

\item {\bf Type II beamforming: }Beamforming is performed by using Type II PMI \cite{3GPP2}. CSI reporting configurations are adjusted for $B=76$. 
\end{itemize}
\begin{figure}[t]
	\centering
  \includegraphics[width=0.55\textwidth]{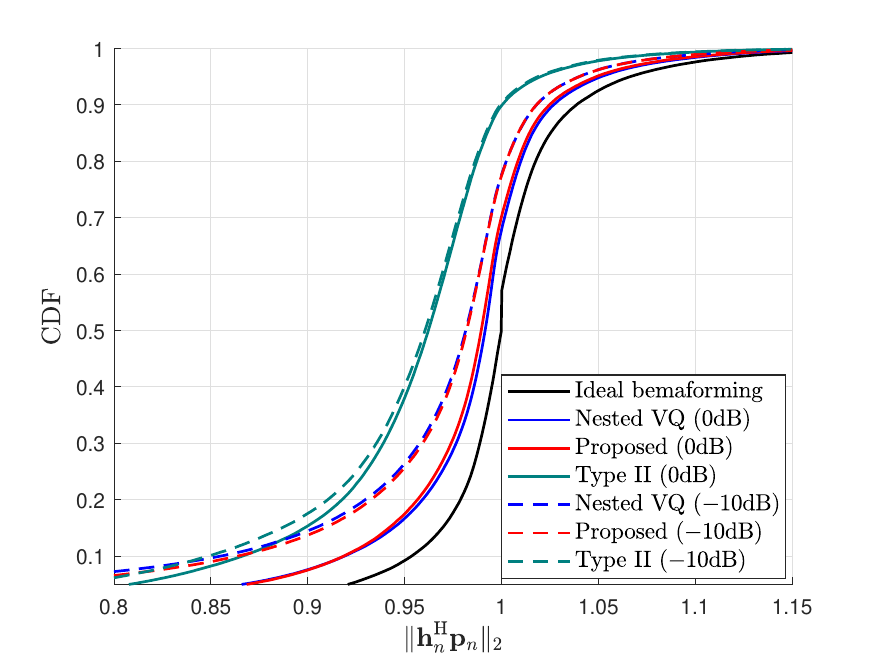}
  \caption{CDFs on precoded NLOS channel gain using various CSI feedback methods with target feedback length $B=76$.} \label{Fig6}
\end{figure}
Even though the DL-based CSI feedback methods are not explicitly trained for $B=76$, they still achieve higher beamforming gains than PMI feedback at both SNR levels. This advantage stems from their ability to efficiently capture the channel's spatial directivity and frequency selectivity (see Fig. 11). For example, in the spatial domain, Type II PMI fails to represent the subtle variations in channel gains, whereas autoencoders precisely capture the differences. Moreover, while PMI imposes a block-segmented structure on frequency selectivity, autoencoders capture it in a continuous form. At high SNR, beamforming gains achieved using VQ slightly exceed those of the proposed method, but, performance inversion occurs at low SNR, where the proposed method demonstrates robustness to channel estimation errors.

%
%

 \begin{figure}[t]
	\centering
  \includegraphics[width=0.85\textwidth]{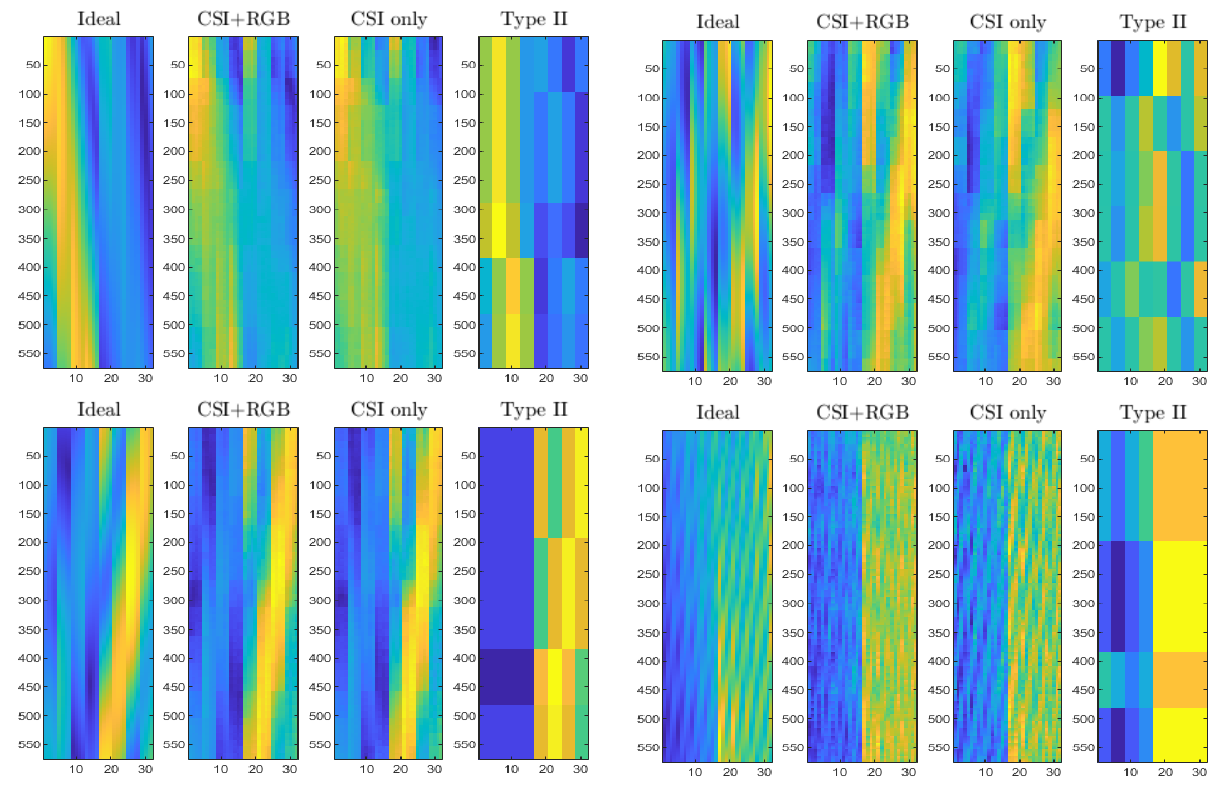}
  \caption{Beamforming matrices using various channel reconstruction methods at $\text{SNR}=-15$dB.} \label{Fig4}
\end{figure}

\subsection{Multi-Modal Channel Reconstruction Using RGB Images}
We consider a scenario, where the BS performs channel reconstruction using computer vision-assisted mode. We use the autoencoder in Fig. 9 as a baseline model and fine tune the channel-relevant feature extraction network $g_{\text E}$ and the refinement network $\mathcal{R}$ (see Table III). We split the RGB image-wireless coupled dataset into 80,000 training data, 32,000 validation data and 16,000 test data. We resize raw images to have a resolution $N_{\text H}\times N_{\text W} = 192\times 256$. The patch embedding sizes are set as $n_{\text H}\times n_{\text W} = 3\times 4$, producing $M=64N_{\text e}$ dimensional feature vector. Under this setup, we train two models with embedding dimensions $[N_{\ell1},N_{\ell2},N_{\ell3},N_{\text e}] = [72, 72, 72, 4]$ and $[72, 96, 128, 32]$. The first model ($N_{\text e}=N_{\rm p}=4$) and the second model ($N_{\text e}=N_{\rm L3}=32$) performs multi-modal processing after the quantization and the linear projection, respectively. The training batch size is $N_{\text{Batch}}=100$ and the network is fine tuned for $300$ epochs. We use the Adam optimizer with $3\times10^{-4}$ learning rate. The target rates for multi-modal fusion are $B=[48, 72, 96, 120, 144]$. Because the RGB image–wireless dataset focuses on LOS positions for the MT, lower SNR levels are used relative to those in the wireless-only channel reconstruction experiments.

\begin{table}[h]
\caption{RGB image fusion model description}
\centering
\renewcommand{\arraystretch}{1.1}
\begin{tabular}{|c|c|c|}
  \hline
  \multicolumn{1}{|c|}{} & \multicolumn{2}{c|}{Number of parameters}  \\ \cline{2-3}
  \multicolumn{1}{|c|}{} & Feature extraction & Refinement   \\ \hline
  CSI + RGB $(N_{\text e}=4)$& 489,004 & 1,716  \\ \hline
 CSI + RGB $(N_{\text e}=32)$& 625,480 & 28,764  \\ \hline
\end{tabular}
\end{table}

\begin{figure}[t]
	\centering
  \includegraphics[width=0.65\textwidth]{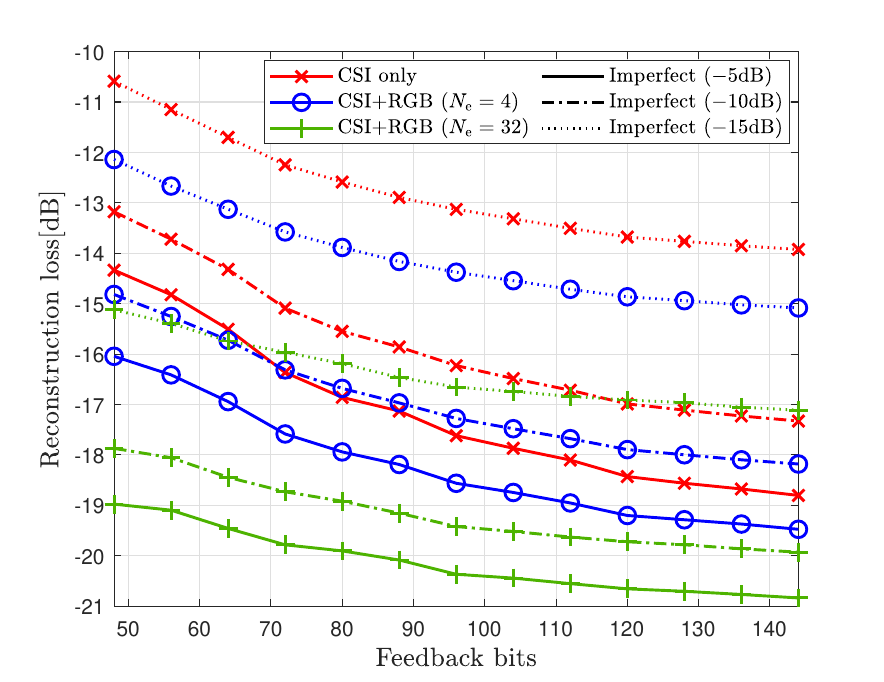}
  \caption{Channel reconstruction loss with and without using RGB images at various feedback rates.} \label{Fig6}
\end{figure}

Fig. 12 compares channel reconstruction losses with (CSI$+$RGB) and without (CSI only) the inclusion of image data at the BS. As anticipated, incorporating larger image features ($N_{\rm e}$) enhances channel reconstruction accuracy. Although the multi-modal fusion network is trained for a limited range of feedback rates, leveraging RGB images improves reconstruction accuracy across all feedback rates. Notably, the performance gains from using image data are more pronounced at lower feedback rates. For example, with perfect CSI input, multi-modal fusion achieves a reconstruction loss improvement of 4.7dB at $B=48$, compared to 2dB at $B=144$. This underscores the network’s ability to dynamically balance CSI and image data contributions based on the feedback rate. Additionally, the performance improvement is even more significant under severe input noise. At $B=96$, using RGB images boosts performance by 3.6dB at SNR $=-15$dB and 2.6dB at SNR $= -5$dB, demonstrating that image information can effectively compensate for channel estimation errors.

Fig. 13 compares the CDFs of beamforming gains at $\text{SNR}=-15$dB. The results indicate that incorporating RGB images into the reconstruction process yields a substantial improvement in beamforming gains. Even under low rate feedback and low SNR conditions, using a large image feature dimension ($N_{\rm e}=32$) can offer near-optimal beamforming gains.

\begin{figure}[t]
	\centering
  \includegraphics[width=0.55\textwidth]{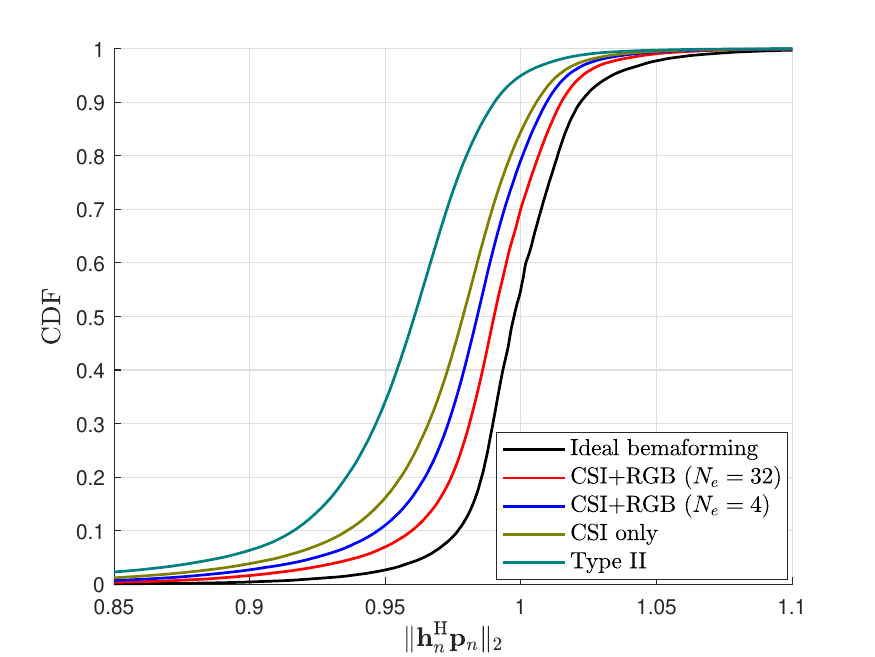}
  \caption{CDFs on precoded LOS channel gain of various CSI feedback methods at $\text{SNR}=-15$dB.} \label{Fig6}
\end{figure}

\begin{figure}[t]
	\centering
  \includegraphics[width=0.95\textwidth]{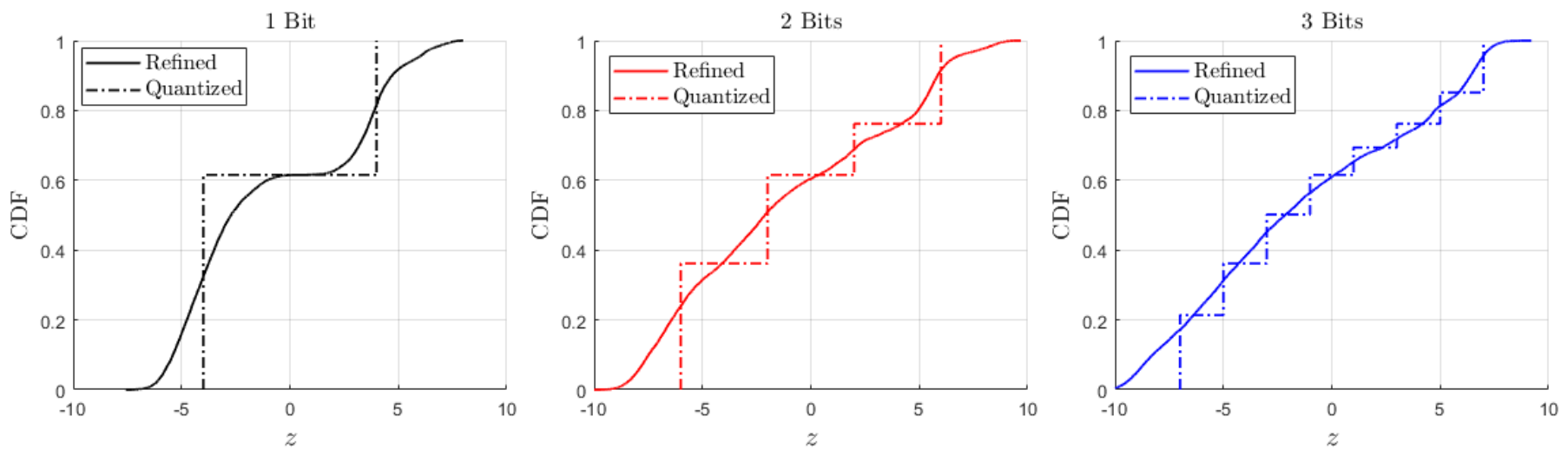}
  \caption{CDFs of quantized feature ${\bf z}({\bf s})$ and refined feature ${\bf z}_{\text r}$ at various feedback rates.} \label{Fig6}
\end{figure}

Fig. 14 compares the CDFs of the quantized feature $z({\bf s})$ and the refined feature $z_{\text r}$ ($N_{\rm e}=4$). The CDFs are evaluated at $B=[48,96,144]$, corresponding to element-wise quantization using 1 to 3 bits. These results provide insight into how RGB images enhance CSI reconstruction accuracy across varying feedback rates. In particular, the multi-modal refinement restores the quantization distortion without overwhelming the foundational discrete feature representation. As the quantization resolution increases, the refined features exhibit progressively smoother distributions. This finding shows that our transfer learning strategy achieves two primary objectives: (1) site-independent CSI reconstruction for wireless-only scenarios, and (2) site-dependent, super-resolution CSI reconstruction in the presence of auxiliary sensor data.

\subsection{Multi-Modal Channel Reconstruction Using Uplink CSI}
We consider a scenario where the BS uses uplink CSI for super-resolution channel reconstruction. In NLOS situations, the BS cannot gather RGB images that capture the MTs. Instead, uplink CSI can be readily obtained via reference signals (e.g., sounding reference signals \cite{3GPP2}). This allows the refinement network to capitalize on the geometrical reciprocity between uplink and downlink channels. Unlike RGB images, uplink CSI is inherently affected by noise. Nonetheless, uplink channel estimation is typically more accurate than its downlink counterpart due to the BS’s large antenna array. In our simulations, we set the uplink channel size the same as the downlink channel. We produce $M=72N_{\text e}$ dimensional uplink channel feature vector through $g_{\text E}(\cdot)$ (see Table IV). We use the same training configurations as in Fig. 12.

\begin{table}[h]
\caption{Uplink CSI fusion model description}
\centering
\renewcommand{\arraystretch}{1.1}
\begin{tabular}{|c|c|c|}
  \hline
  \multicolumn{1}{|c|}{} & \multicolumn{2}{c|}{Number of parameters}  \\ \cline{2-3}
  \multicolumn{1}{|c|}{} & Feature extraction & Refinement   \\ \hline
  CSI + Uplink CSI $(N_{\text e}=4)$ & 506,988 & 1,844  \\ \hline
  CSI + Uplink CSI $(N_{\text e}=32)$ & 578,772 & 29,116  \\ \hline
\end{tabular}
\end{table}

\begin{figure}[t]
	\centering
  \includegraphics[width=0.65\textwidth]{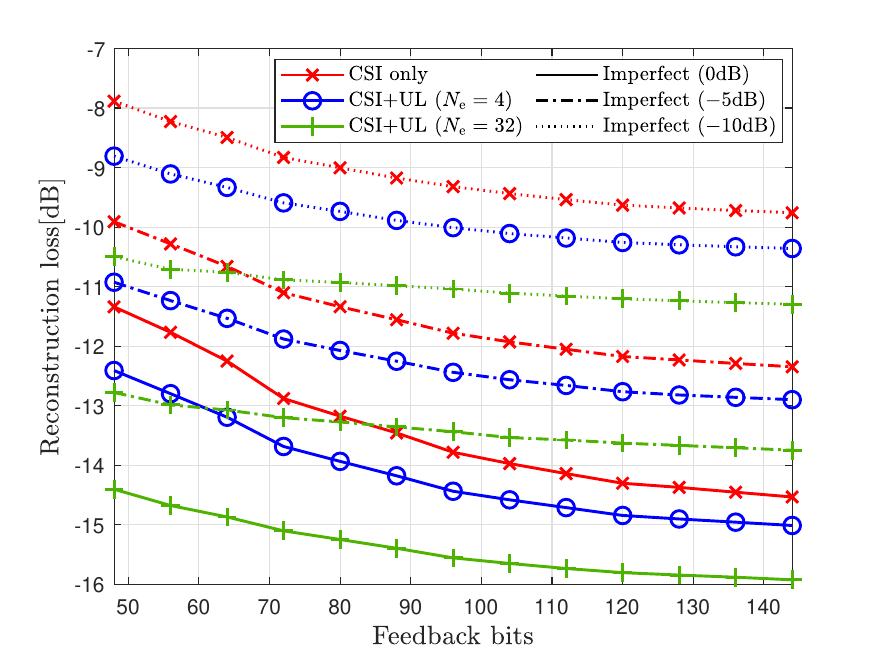}
  \caption{Channel reconstruction loss with and without using uplink CSI fusion at various feedback rates.} \label{Fig6}
\end{figure}

Fig. 15 compares the channel reconstruction losses with (CSI$+$UL) and without (CSI only) uplink CSI. Despite its inherent noise, uplink CSI follows a trend similar to RGB images when employed for multi-modal channel reconstruction. For instance, integrating uplink CSI improves reconstruction accuracy at all feedback rates, with notable gains at lower rates. However, the performance boost due to uplink CSI does not grow as SNR decreases. For example, at $B=192$, $N_{\rm e}=4$ and $N_{\rm e}=32$ yield consistent gains of 0.5dB and 1.4dB  across all SNRs, respectively. This reflects the noise that degrades the accuracy of both downlink and uplink CSI.

\begin{figure}[t]
	\centering
  \includegraphics[width=0.55\textwidth]{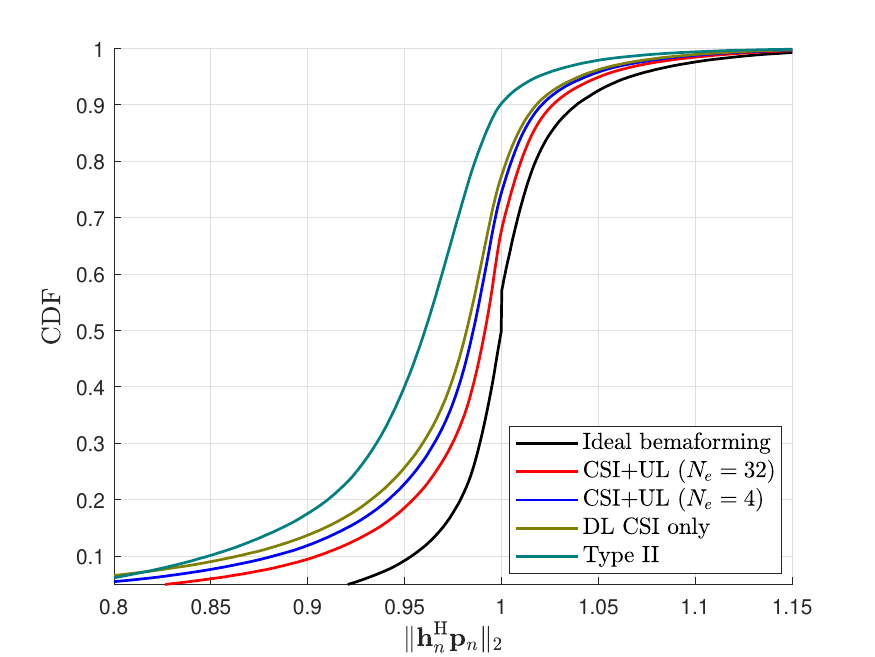}
  \caption{CDFs on precoded NLOS channel gain of various CSI feedback methods at $\text{SNR}=-10$dB.} \label{Fig6}
\end{figure}

Fig. 16 compares the CDFs of beamforming gains at $\text{SNR}=-10$dB. Even under low SNR condition, incorporating uplink CSI leads to a notable improvement in beamforming gains. However, in contrast to the RGB image-fusion (see Fig. 13), the worst-case beamforming gains exhibit a large gap from the optimal gains. This shows that the flawed labels are limited in restoring CSI, particularly when noisy CSI significantly deviates from the true ones.

\section{Conclusion}
We have proposed a multi-modal CSI reconstruction technique for FDD massive MIMO downlink communications. The key innovation of this framework is to exploit the inherent correlation between feedback CSI and sensor data sourced from the same physical environment. By leveraging multi-modal fusion, the hybrid approach overcomes the limitations of both wireless-only and computer vision-only methods. In particular, high resolution sensor data mitigates the feedback overhead bottleneck of wireless mechanisms by compensating for CSI distortions due to noise, compression, and quantization. Conversely, wireless data allows the BS to support MTs in challenging scenarios, such as NLOS locations or adverse weather conditions. Central to our approach is an autoencoder that produces disjoint quantization outputs at different rates. Using this network, we have demonstrated that super-resolution CSI reconstruction is possible under diverse CSI reporting configurations. Simulation results reveal that employing RGB images or uplink CSI for super-resolution CSI reconstruction can achieve near-optimal beamforming gains in 5G NR-compliant scenarios. A promising direction for future research is to extend this framework to incorporate a broader range of sensing modalities and to explore scenarios where sensors are not co-located with the BS.



\begin{thebibliography}{1}

\bibitem{Swindlehurst}
A. L. Swindlehurst, E. Ayanoglu, P. Heydari, and F. Capolino,  ``Millimeter-wave massive MIMO: the next wireless revolution?," {\em IEEE Commun. Mag.,} vol. 52, no. 9, pp. 56-62, Sep. 2014.


\bibitem{Boccardi}
F. Boccardi, R. W. Heath Jr., A. Lozano, T. L. Marzetta, and P. Popovski, ``Five disruptive technology directions for 5G," {\em IEEE Commun. Mag.,} vol. 52, no. 2, pp. 74-80, Feb. 2014.

\bibitem{Larsson}
E. G. Larsson, O. Edfors, F. Tufvesson, and T. L. Marzetta,  ``Massive MIMO for next generation wireless systems," {\em IEEE Commun. Mag.,} vol. 52, no. 2, pp. 186-195, Feb. 2014.

\bibitem{Ayach}
O. E. Ayach, S. Rajagopal, S. Abu-Surra, Z. Pi, and R. W. Heath Jr., ``Spatially sparse precoding in millimeter wave MIMO systems," {\em IEEE Trans. Wireless Commun.,} vol. 13, no. 3, pp. 1499-1513, Mar. 2014.

\bibitem{Han}
S. Han, C.-L. I, Z. Xu, and C. Rowell, ``Large-scale antenna systems with hybrid analog and digital beamforming for millimeter wave 5G," {\em IEEE Commun. Mag.,} vol. 53, no. 1, pp. 186-194, Jan. 2015.

\bibitem{Love}
D. J. Love, R. W. Heath Jr., V. K. N. Lau, D. Gesbert, B. D. Rao, and M. Andrews, ``An overview of limited feedback in wireless communication systems, {\em IEEE J. Sel. Areas Commun.,} vol. 26, no. 8, pp. 1341-1365, Oct. 2008.


\bibitem{Guo2}
J. Guo, C.-K. Wen, S. Jin, and G. Y. Li, ``Overview of deep learning-based CSI feedback in massive MIMO systems," {\em IEEE Trans. Commun.,} vol. 70, no. 12, pp. 8017-8045, Dec. 2022.


\bibitem{Shen}
W. Shen, L. Dai, B. Shim, Z. Wang, and R. W. Heath Jr., ``Channel feedback based on AoD-adaptive subspace codebook in FDD massive MIMO systems," {\em IEEE Trans. Commun.,} vol. 66, no. 11, pp. 5235-5248, Nov. 2018.

\bibitem{Ju1}
H. Ju, S. Jeong, B. Lee, and B. Shim, `Transformer-assisted parametric CSI feedback for mmWave massive MIMO systems," {\em IEEE Trans. Wireless Commun.,} vol. 23, no. 12, pp. 18774-18787, Dec. 2024.

\bibitem{3GPP1}
3GPP, ``Technical specification group radio access network, physical layer procedures for data (Release 18)," TS 38.214 V18.5.0, Dec. 2024.

\bibitem{Qin}
Z. Qin and H. Yin, ``A review of codebooks for CSI feedback in 5G new radio and beyond," {\em arXiv preprint arXiv:2302.09222,} 2023.


\bibitem{Eltayeb}
M. E. Eltayeb, T. Y. Al-Naffouri, and H. R. Bahrami, ``Compressive sensing for feedback reduction in MIMO broadcast channels," {\em IEEE Trans. Commun.,} vol. 62, no. 9, pp. 3209-3222, Sep. 2014.

\bibitem{Huang}
X.-L. Huang, J. Wu, Y. Wen, F. Hu, Y. Wang, and T. Jiang, ``Rate-adaptive feedback with Bayesian compressive sensing in multiuser MIMO beamforming systems," {\em IEEE Trans. Wireless Commun.,} vol. 15, no. 7, pp. 4839-4851, July 2016.

\bibitem{Gao}
Z. Gao, L. Di, S. Han, C.-L. I, Z. Wang, and L. Hanzo, ``Compressive sensing techniques for next-generation wireless communications," {\em IEEE Wireless Commun.,} vol. 25, no. 3, pp. 144-153, June 2018.

\bibitem{Kulsoom}
F. Kulsoom, A. Vizziello, H. N. Chaudhry, and P. Savazzi, ``Joint sparse channel recovery with quantized feedback for multi-user massive MIMO systems," {\em IEEE Access,} vol. 8, pp. 11046-11060, Jan. 2020.

\bibitem{3GPPAI}
3GPP, “Technical specification group radio access network, study on artificial intelligence (AI)/machine learning (ML) for
NR air interface (Release 18),” TR 38.843, V18.0.0, Dec. 2023.

\bibitem{Liang}
P. Liang, J. Fan, W. Shen, Z. Qin, and G. Y. Li, ``Deep learning and compressive sensing-based CSI feedback in FDD massive MIMO systems," {\em IEEE Trans. Veh. Tech.,} vol. 69, no. 8, pp. 9217-9222, Aug. 2020.


\bibitem{Wen}
C.-K. Wen, W.-T. Shih, and S. Jin, ``Deep learning for massive MIMO CSI feedback," {\em IEEE Wireless Commun. Letters,} vol. 7, no. 5, pp. 748-751, Oct. 2018.

\bibitem{Mashhadi}
M. B. Mashhadi, Q. Yang, and D. Gündüz, ``Distributed deep convolutional compression for massive MIMO CSI feedback," {\em IEEE Trans. Wireless Commun.,} vol. 20, no. 4, pp. 2621-2633, Apr. 2021.

\bibitem{Lu}
C. Lu, W. Xu, H. Shen, J. Zhu, and K. Wang, ``MIMO channel information feedback using deep recurrent network," {\em IEEE Commun. Letters,} vol. 23, no. 1, pp. 188-191, Jan. 2019. 

\bibitem{Liu}
Z. Liu, L. Zhang, and Z. Ding, ``An efficient deep learning framework for low rate massive MIMO CSI reporting," {\em IEEE Trans. Commun.,} vol. 68, no. 8, pp. 4761-4772, Aug. 2020.

\bibitem{Zeng}
X. Zhang, Z. Lu, R. Zeng, and J. Wang, ``Quantization adaptor for bit-level deep learning-based massive MIMO CSI feedback," {\em IEEE Trans. Veh. Tech.,} vol. 73, no. 4, pp. 5443-5453, Apr. 2024.





\bibitem{Guo}
J. Guo, C.-K. Wen, S. Jin, and G. Y. Li, ``Convolutional neural network-based multiple-rate compressive sensing for massive MIMO CSI feedback: design, simulation, and analysis," {\em IEEE Trans. Wireless Commun.,} vol. 19, no. 4, pp. 2827-2840, Apr. 2020.

\bibitem{Liang2}
X. Liang, H. Chang, H. Li, X. Gu and L. Zhang, ``Changeable rate and novel quantization for CSI feedback based on deep learning," {\em IEEE Trans. Wireless Commun.,} vol. 21, no. 12, pp. 10100-10114, Dec. 2022.


\bibitem{Jeon1}
J. Shin, Y. Kang, and Y.-S. Jeon, ``Vector quatization for deep-learning-based CSI feedback in massive MIMO systems," {\em IEEE Wireless Commun. Letters,} vol. 13, no. 9, pp. 2382-2376, Sep. 2024.



\bibitem{Inf}
T. M. Cover and J. A. Thomas, {\em Elements of Information Theory,}  New york, NY, USA: Wiley, 2006.


\bibitem{FLiu}
F. Liu, Y. Cui, C. Masouros, J. Xu, T. X. Han, Y. C. Eldar, and S. BuzziM ``Integrated sensing and communications: toward dual-function wireless networks for 6G and beyond," {\em IEEE J. Sel. Areas Commun.,} vol. 40, no. 6, pp. 1728-1767, June 2022.

\bibitem{Zheng}
K. Zheng, H. Yang, Z. Ying, P. Wang, and L. Hanzo, ``Vision-assisted millimeter-wave beam management for next-generation wireless systems: concepts, solutions, and open challenges,"  {\em IEEE Veh. Tech. Mag.,} vol. 18, no. 3, pp. 58-68, Sep. 2023.

\bibitem{Kim}
S. Kim, J. Moon, J. Wu, B. Shim, and M. Z. Win, ``Vision-aided positioning and beam focusing for 6G terahertz communications," {\em IEEE J. Sel. Areas Commun.,} vol. 42, no.9, pp. 2503-2519, Sep. 2024.


\bibitem{Ahn}
Y. Ahn, J. Kim, S. Kim, S. Kim, and B. Shim, ``Sensing and computer vision-aided mobility management for 6G millimeter and terahertz communication systems," {\em IEEE Trans. Commun.,} vol. 72, no. 10, pp. 6044-6058, Oct. 2024.


\bibitem{Ahn2} 
Y. Ahn, J. Kin, S. Kim, K. Shim, J. Kim, S. Kim, and B. Shim, ``Toward intelligent millimeter and terahertz communication for 6G: computer vision-aided beamforming," {\em IEEE Wireless Commun.,} vol. 30, no. 5, pp. 179-186, Oct. 2023.


\bibitem{Shimomura}
H. Shimomura, Y. Koda, T. Kanda, K. Yamamoto, T. Nishio, and A. Taya, ``Vision-aided frame-capture-based CSI recomposition for WiFi sensing: a multi modal approach," {\em IEEE Consumer Commun. Net. Conf.,} 2023.

\bibitem{ViWi}
M. Alrabeiah, A. Hredzak, Z. Liu, and A. Alkhateeb, ``ViWi: a deep learning dataset framework for vision-aided wireless communications," {\em IEEE Veh. Tech. Conf.,} 2020. 



\bibitem{3GPP2}
3GPP, ``Technical specification group radio access network, physical channels and modulation (Release 18)," TS 38.211 V18.5.0, Dec. 2024.


\bibitem{3GPP3}
3GPP, ``Technical specification group radio access network, study on channel model for frequencies from 0.5 to 100GHz (Release 18)," TR 38.901 V18.0.0, Mar. 2024.


\bibitem{VQ-VAE}
A. V. D. Oord, O. Vinyals, and K. Kavukcuoglu, ``Neural discrete representation learning," {\em Adv. Neural Inf. Process. Syst.,} 2017.


\bibitem{Swin}
Z. Liu, Y. Lin, Y. Cao, H. Hu, Y. Wei, Z. Zhang, S. Lin, and B. Guo, ``Swin Transformer: Hierarchical vision Transformer using shifted windows," {\em IEEE Int. Conf. Comp. Vision,} 2021.


\bibitem{MM}
P. Xu, X. Zhu, and D. A. Clifton, ``Multimodal learning with Transformers: a survey," {\em arXiv preprint arXiv:2206.06488,} 2022.



\bibitem{Ahmed}
G. Charan and A. Alkhateeb, ``User identification: a key enabler for multi-user vision-aided communications," {\em IEEE Open Journal Commun. Society,} vol. 5, pp. 472-488, Dec. 2023.


\bibitem{Trans}
A. Vaswani, N. Shazeer, N. Parmar, J. Uszkoreit, L. Jones, A. N. Gomez, L. Kaiser, and I. Polosukhin, ``Attention is all you need," {\em Adv. Neural Inf. Process. Syst.,} 2017.



\bibitem{LBAE}
J. Fajtl, V. Argyriou, D. Monekosso, and P. Remagnino, ``Latent Bernoulli autoencoder," {\em Int. Conf. Machine Learn.,} 2020.






 
\end{thebibliography}
\end{document}